\documentclass[]{interact}

\RequirePackage{amsthm,amsmath,amsfonts,amssymb}
\RequirePackage[colorlinks,citecolor=blue,urlcolor=blue]{hyperref}
\RequirePackage{graphicx}
\RequirePackage{dsfont,enumerate,multirow,rotating}

\usepackage[caption=false]{subfig}

\usepackage[numbers,sort&compress]{natbib}
\bibpunct[, ]{[}{]}{,}{n}{,}{,}
\makeatletter
\def\NAT@def@citea{\def\@citea{\NAT@separator}}
\makeatother

\newcommand{\1}{\mathds{1}}

\newcommand{\diag}{\text{diag}}			   	%
\newcommand{\fcond}{\boldsymbol{[\cdots]}} 	%
\newcommand{\mA}{\boldsymbol{A}} %
\newcommand{\mB}{\boldsymbol{B}} %
\newcommand{\mF}{\boldsymbol{\mathcal{F}}} %
\newcommand{\mG}{\boldsymbol{G}} %
\newcommand{\mH}{\boldsymbol{H}} %
\newcommand{\mHH}{\boldsymbol{\mathcal{H}}} %
\newcommand{\mI}{\boldsymbol{I}} %
\newcommand{\mJ}{\boldsymbol{J}} %
\newcommand{\mm}{\boldsymbol{\mathcal{M}}} %
\newcommand{\mP}{\boldsymbol{\Phi}} %
\newcommand{\mS}{\boldsymbol{\Sigma}} %
\newcommand{\mV}{\boldsymbol{V}} %
\newcommand{\mW}{\boldsymbol{W}} %
\newcommand{\ones}{\boldsymbol{1}} %
\newcommand{\p}{\mathds{P}}				   	%
\newcommand{\R}{\mathds{R}}                	%

\newcommand{\T}{'}					   	%
\newcommand{\TT}{\mathcal{T}}			%
\newcommand{\vb}{\boldsymbol{\beta}} %
\newcommand{\va}{\boldsymbol{\alpha}} %
\newcommand{\valpha}{\boldsymbol{\alpha}} %
\newcommand{\ve}{\boldsymbol{e}} %
\newcommand{\veps}{\boldsymbol{\epsilon}} %
\newcommand{\vF}{\boldsymbol{F}} %
\newcommand{\vmu}{\boldsymbol{\mu}} %
\newcommand{\vphi}{\boldsymbol{\phi}} %
\newcommand{\vomega}{\boldsymbol{\omega}} %
\newcommand{\vt}{\boldsymbol{\theta}} %
\newcommand{\vv}{\boldsymbol{v}} %
\newcommand{\vvt}{\boldsymbol{\vartheta}} %
\newcommand{\vx}{\boldsymbol{x}} %
\newcommand{\vX}{\boldsymbol{X}} %
\newcommand{\vy}{\boldsymbol{y}} %
\newcommand{\vz}{\boldsymbol{z}} %
\newcommand{\vect}[1]{\boldsymbol{#1}}      %
\newcommand{\zeroes}{\boldsymbol{0}} %

\newcommand{\rv}{r.v.}                	%
\newcommand{\bern}{{Bern}}    	%
\newcommand{\logit}{\text{logit}}    	%
\newcommand{\ch}[1]{\left\{ #1 \right\}}   	%
\newcommand{\co}[1]{\left[ #1 \right]}	   	%
\newcommand{\pa}[1]{\left( #1 \right)}	   	%
\newcommand{\comment}[1]{}			   	%

\theoremstyle{plain}
\newtheorem{theorem}{Theorem}[section]

\newtheorem{proposition}[theorem]{Proposition}

\theoremstyle{definition}

\theoremstyle{remark}

\begin{document}


\title{Bayesian estimation of dynamic weights in Gaussian mixture models}

\author{
\name{Michel H. Montoril\textsuperscript{a}\thanks{CONTACT M.~H. Montoril. Email: michel@ufscar.br}; Leandro T. Correia\textsuperscript{b} and Helio S. Migon\textsuperscript{c}}
\affil{\textsuperscript{a}Department of Statistics,
Federal University of S\~ao Carlos, S\~ao Carlos, Brazil; \\ \textsuperscript{b}Department of Statistics,
University of Bras\'ilia, Bras\'ilia, Brazil; \\
\textsuperscript{c}Department of Statistics,
Federal University of Rio de Janeiro, Rio de Janeiro, Brazil}
}

\maketitle

\begin{abstract}
This paper proposes a generalization of Gaussian mixture models, where the mixture weight is allowed to behave as an unknown function of time. This model is capable of successfully capturing the features of the data, as demonstrated by simulated and real datasets. It can be useful in studies such as clustering, change-point and process control. In order to estimate the mixture weight function, we propose two new Bayesian nonlinear dynamic approaches for polynomial models, that can be extended to other problems involving polynomial nonlinear dynamic models. One of the methods, called here component-wise Metropolis-Hastings, apply the Metropolis-Hastings algorithm to each local level component of the state equation. It is more general and can be used in any situation where the observation and state equations are nonlinearly connected. The other method tends to be faster, but is applied specifically to binary data (using the probit link function). The performance of these methods of estimation, in the context of the proposed dynamic Gaussian mixture model, is evaluated through simulated datasets. Also, an application to an array Comparative Genomic Hybridization (aCGH) dataset from glioblastoma cancer illustrates our proposal, highlighting the ability of the method to detect chromosome aberrations.
\end{abstract}


\section{Introduction}

Gaussian mixture models (GMM) have been used to solve problems in a wide range of fields, under different scenarios. In the context of statistical learning, these models play an important role. We can highlight clustering \citep{Saraiva-Milan-2012} and classification \citep{Fernando-et-al-2012} as unsupervised and supervised learning examples. For more details and examples, see \cite{HastieTibshiraniFriedman2017}.

The important role played by GMMs makes them topics of interest of various researchers, who have improved and generalized this class in the context of (homogeneous) hidden Markov models (HMM). Examples range from econometrics \citep{Ryden-Terasvirta-Asbrink-1998} to genetics \citep{Boys-Henderson-2004}. In the multivariate case, \cite{Spezia-2010} introduced Gaussian hidden Markov models with unknown number of regimes.

Still in terms of generalization, we can mention the non-homogeneous HMM (NHMM), where the transition probabilities are not constant. For example, \cite{Raymond-Rich-1997} considered binary probit models to link covariates to the transition probabilities; \cite{Meligkotsidou-Dellaportas-2011} developed a Bayesian forecasting method where the transition probabilities depend on covariates; and \cite{Holsclaw-et-al-2017} developed an efficient MCMC sampling scheme. In the spirit of Bayesian non-parametrics, another generalization is the infinite HMM (iHMM), where the HMM is allowed to have a countably infinite number of hidden states \citep{Beal.etal-2002, Teh.etal-2006-JASA}.

In this work we analyze a typical GMM, but using a state-space (SS) approach to model the time evolution of the mixture weights. The ``non-homogeneous'' structure of the model leads to think of it as a type of  NHMM. Both classes, SS and (N)HMM, are similar in the sense that they relate unobserved states to responses. In SS, the states are continuous, while in (N)HMM they are discrete \citep{Fahrmeir-Tutz-2001-chap-8}. Specifically comparing our approach with NHMM, we can highlight that the latter focus on the components of the mixture as unobserved states, dealing with transition probabilities that vary (non-homogeneous) along time. With respect to the methodology proposed in this paper, the unobserved states are the (dynamic) mixture weights. In other words, we assume directly a dynamic behavior for the mixture weights and deal with them using some smoothing method (in this work we consider an SS approach for such a task). This argument makes our model flexible and better able to capture the data features. Therefore, despite the similarities of dealing with analogous problems, these methodologies are not exactly comparable, in the sense that transition probabilities (NHMM) and dynamic mixture weights (our proposal) are different things. 

The model as formulated in this paper, although sophisticated, is simple and allows for classification, clustering, change-points detection and process control. The Bayesian nature of the procedure of estimation provides conditions to estimate both, the component parameters and the dynamic mixture weights. This generalizes a similar model studied by \cite{MontorilPinheiroVidakovic2019}, where the authors considered the mixture of two random variables (\rv's), assuming known means and variances, with unknown time-varying mixture weights (estimated by wavelet bases). Another similar model was used by \cite{Scaccia.Green-2003-JCGS} to study growth curves, where the authors considered non-parametric weights estimated by cubic B-splines.

To the best of our knowledge, despite the similarities above mentioned, the model considered here is a novelty, as well as the method of estimation, that modifies the precision-based algorithm studied by \cite{Chan-Jeliazkov-2009}. Basically, we exploit the Markovian structure of polynomial dynamic linear models by reordering the state vector. Such a change provides an efficient algorithm that is able to estimate the state vector in a single step. Furthermore, based on this modification, we generalize the studies to the case of nonlinear dynamic models, which is applicable to the estimation of the dynamic mixture weights, as well as the dynamic generalized linear models \citep{West-Harrison-Migon-1985}. Two efficient and scalable simulation algorithms are proposed: one general, that performs component-wise Metropolis-Hastings; and another more efficient, but specific to Bernoulli data, that consider the properties of probit link with binary data \citep{Albert-Chib-1993}. The idea of estimating in one step results in efficiency improvements \citep{Carter-Kohn-1994, Fruhwirth-Schnatter-1994}. This motivated us to propose the new algorithms above mentioned.

This paper is organized as follows. In Section \ref{sec:dinmixmodel}, a proposed Bayesian method for linear and nonlinear dynamic models is discussed. This method is employed for estimation of the dynamic mixture weights. In Section \ref{sec:mixmodel},  the dynamic Gaussian mixture model is introduced and its inference is discussed. The performance of the method proposed in Section \ref{sec:dinmixmodel} is evaluated in Section \ref{sec:simul-data} by using simulated datasets, where four different functional behaviors are considered as dynamic weights for Gaussian mixture data. In Section \ref{sec:application}, we apply the method to an array Comparative Genomic Hybridization (aCGH) dataset from glioblastoma cancer studies. Some concluding remarks are given in Section \ref{sec:comments}.

\section{Adapting the precision-based method for polynomial models} \label{sec:dinmixmodel}

In this section, we discuss the inference for polynomial dynamic models, following  an alternative route than the usual FFBS \citep{Fruhwirth-Schnatter-1994}, where we consider the precision-based method by \cite{Chan-Jeliazkov-2009}. The polynomial model structure is explored to make some improvements in the Chan's algorithm in the Gaussian scenario, where we apply a simple reordering of the state vector for such an improvement. This initial study of Gaussian dynamic linear models, besides the improvement, is useful to facilitate comprehension of the proposal and specify the notation. Then a generalization to the nonlinear situation is explored, where we propose two methods of estimation: one general, that is able to deal with different situations of nonlinearity, and another more specific, for Bernoulli data. These nonlinear methods, obtained based on the polynomial structure of the dynamic models, will be important in the estimation process of the dynamic mixture weights, discussed in Section \ref{sec:dynmixw}.

\subsection{Bayesian inference in dynamic Gaussian polynomial models}\label{sec:idgp}

Following \cite{WestHarrison1997}, we define a dynamic linear model (DLM) by the quadruple $\{\vF, \mG, V, \mW\}_t$, where $\vF_t$ is a known vector of constants or predictor variables (features or regressors), $\mG_t$ is a known state vector ($\vt_t$) evolution matrix, $\mW_t$  is the variance of the stochastic evolution innovation vector, and $V_t$ is the observational variance. Without loss of generality, we assume that $\mW_t = \diag\{W_1, \ldots, W_p\}$, $ \forall t $.  The {$p$-th order polynomial model} is similar to the canonical model: $\vF_t = (1,0, \cdots, 0)^{\T}$, a vector of size $p$, and $ \mG_t = \mJ_p(1)$ (a Jordan block, which has unit eigenvalue with multiplicity $ p $). 

For the sake of information, the second order polynomial DLM is related to an important non-parametric tool, namely cubic splines \citep{Wahba-1990, Green-Silverman-1993, Eubank-1999}. In \cite{Kohn-Ansley-1987}, the authors write the spline smoothing formulation of \cite{Wahba-1978} as a stochastic difference equation and represent it in the state-space form. Therefore, for equally spaced data, it is easy to see that a cubic spline corresponds to a dynamic model with $ \vF = (1, 0)^{\T} $ and $ \mG = \mJ_2(1) $. This formulation was further extended to generalized additive regression models by \cite{Biller-Fahrmeir-1997}.

To facilitate comprehension of the employed method, we devote this subsection to discuss the dynamic Gaussian models. In a general framework, we consider the model composed by the observation equation and the state evolution equation
\begin{eqnarray}\label{mod:gendln-obs}
y_t & = & \vF_t^{\T}\vt_t + \epsilon_t, \\ \label{mod:gendln-ss}
\vt_t & = & \mG_t \vt_{t-1} + \vomega_t,
\end{eqnarray}
where $\epsilon_t \sim N(0, V_t)$ and $\vomega_t \sim N(\zeroes, \mW_t)$, $t = 1, 2, \ldots, T$.

Most of the literature involving state-space dynamic models deals with the Kalman filtering and smoothing recursions, in order to obtain the joint posterior distributions of the states \citep[for more details, see e.g.][]{WestHarrison1997, Kroese-Chan-2014}. This tends to be computationally intensive, so joint sampling directly from $ p(\vt | \text{ data}) $ is more efficient \citep{Carter-Kohn-1994, Fruhwirth-Schnatter-1994}. 

In this work we consider a \textit{precision-based algorithm}, a sampling method to obtain the latent states in a single step, avoiding the two steps procedure used by \cite{Carter-Kohn-1994} and \cite{Fruhwirth-Schnatter-1994}, and exploiting the sparse feature of normal precision matrices (which justifies the name initially employed by \cite{Chan-Strachan-2012} and used adopted in this paper). This is a scalable procedure and greatly facilitates subsequent aspects of the analysis. This precision-based algorithm has been successfully used by Joshua Chan and collaborators to solve different problems. A few references include some recent papers, like \cite{Chan-Koop-Potter-2013}, \cite{Chan-Eisenstat-2018}, \cite{Chan-Eisenstat-Strachan-2020} and \cite{Zhang.etal-2020}. A seminal work is \cite{Chan-Jeliazkov-2009}, while interesting and didactic discussions can be found in \cite{Chan-Strachan-2012} and \cite{Kroese-Chan-2014}.

In a brief explanation, one can show that the model in \eqref{mod:gendln-obs}-\eqref{mod:gendln-ss} can be rewritten as
\begin{eqnarray}
\vy | \vt, \mV & \sim & N(\mF\vt, \mV), \label{mod:mat-genln-obs} \\ 
\vt| \vt_0, \mW & \sim & N\co{\mHH^{-1} \mm \vt_0, \pa{\mHH^{\T} \mW^{-1} \mHH}^{-1}}, \label{mod:mat-gendln-ss}
\end{eqnarray}
which corresponds to a simple Bayesian regression model, where $ \vy = (y_1, \ldots, y_T)^{\T} $, $\mF = \diag\{\vF_1', \ldots, \vF_T'\}$, $\vt = (\vt_1^{\T}, \ldots, \vt_T^{\T})^{\T}$, $\mV = \diag\{V_1, \ldots, V_T\}$, $\mW = \diag\{\mW_1, \ldots, \mW_T\}$, $ \mm = \pa{\mG_1^{\T}, \zeroes^{\T}, \ldots, \zeroes^{\T}}^{\T} $ and $ \mHH $ is a block matrix of the form $\mHH_{ij} = \mI $,  if $ i = j, i = 1, \ldots, T$; $- \mG_i $, if $ i = j-1, i = 2, \ldots, T$; and $\zeroes$, otherwise. It should be noted that \eqref{mod:mat-genln-obs} and \eqref{mod:mat-gendln-ss} correspond to a Bayesian regression model with a very sparse precision matrix \citep[for details, see][]{Chan-Strachan-2012}.

Due to conjugation, we can easily derive the posterior 
$
\vt|\vy, \mV, \vt_0, \mW \sim N\pa{\bar{\vmu}, \bar{\mP}^{-1}},
$
where
\begin{eqnarray*}
\bar{\mP} & = & \mHH^{\T} \mW^{-1} \mHH + \mF^{\T} \mV^{-1} \mF, \\
\bar{\vmu} & = & \bar{\mP}^{-1}\left(\mHH^{\T} \mW^{-1} \mm \vt_0 + \mF^{\T} \mV^{-1} \vy\right).
\end{eqnarray*}
One can see that, following the structure of the precision matrix in \eqref{mod:mat-gendln-ss}, the posterior precision matrix is also sparse of the band type. The sparseness of the posterior precision matrix allows easily generating the states in a single step, resulting in better efficiency. For this reason, the method is known as precision-based algorithm. The computational advantages of this approach are discussed in more details by \cite{McCausland-Miller-Pelletier-2011}. Moreover, \cite[Chapter~4]{Golub-VanLoan-2013} discuss the advantages in terms of the number of operations, involving (sparse) band matrices.

\subsubsection{Modifying the precision-based algorithm in the polynomial case} \label{ssec:imp-pol}

Besides the benefits of considering the precision-based algorithm as presented above, depending on the model, it can still be improved. For polynomial DLMs, the following Markovian structure is implied by the Jordan form:
\begin{equation}\label{thet:markov}
p(\vvt) = p(\vvt_1|\vvt_2) \cdots p(\vvt_{p-1}|\vvt_p) p(\vvt_p),
\end{equation}
where $ \vvt_{k} = (\theta_{1k}, \ldots, \theta_{Tk})' $, $ k = 1, \ldots, p $, and $ \vvt = (\vvt_{1}', \ldots, \vvt_{p}')' $. We omit the hyperparameters to avoid overloading the notation. Observe that, by definition, $ \vvt $ is simply a reordering of $ \vt $ in \eqref{mod:mat-gendln-ss}, which corresponds to an orthonormal transformation of $ \vt $. Therefore, all developments presented before can be easily adapted.

When the modeling is based on polynomial DLMs, it is even possible to preserve the banded property of the precision matrix. For the sake of simplicity, we consider the case of a homoscedastic model with independent innovations, i.e., $V_t \equiv V$ and $\mW_t \equiv \mW = \diag\{W_1, \ldots, W_p\}$, $t = 1, 2, \ldots, T$. This kind of simplification, in \eqref{mod:mat-genln-obs}-\eqref{mod:mat-gendln-ss}, does not provide much computational improvement. However, under the proposed reordering, the Markovian property in \eqref{thet:markov} provides an additional simplification of model \eqref{mod:mat-genln-obs}-\eqref{mod:mat-gendln-ss}, which is expressed in the proposition below.
\begin{proposition}\label{prop:prior-obs-states}
Let $ \vvt $ be the state vector ordered as in \eqref{thet:markov}. Denote $ \ones = (1, \ldots, 1)' $, a vector of size $ T $, and $ \mH $, a $ T \times T $ band matrix, which has $ 1 $ in its main diagonal, $ -1 $ in the sub-diagonal and zero elsewhere. Under the assumption of homoscedasticity with independent innovations, the polynomial DLM can be written as:
\begin{eqnarray}
\vy | \vvt_{1} & \sim & N(\vvt_{1}, V \mI), \label{mod:mat-pol-obs}\\
\vvt_k | \vvt_{k+1} & \sim & N\co{\vmu_k, W_k\pa{\mH^{\T}\mH}^{-1}}, k = 1, \ldots, p-1, \label{mod:mat-pol-ss1} \\
\vvt_p & \sim & N\co{\vmu_p, W_p\pa{\mH^{\T}\mH}^{-1}}, \label{mod:mat-pol-ss2}
\end{eqnarray}
where $ \vmu_k = (\theta_{0k} + \theta_{0(k+1)})\ones + (\mH^{-1} - \mI) \vvt_{k+1} $, $k = 1, 2, \ldots, p-1$, and $ \vmu_p = \theta_{0p}\ones $.
\end{proposition}

The model above makes clear the Markovian property \eqref{thet:markov}. Furthermore, the model in \eqref{mod:mat-pol-obs}-\eqref{mod:mat-pol-ss2} is similar to the model in \eqref{mod:mat-genln-obs}-\eqref{mod:mat-gendln-ss}, with the same features of sparseness and band type, representing a simple regression with a special prior. Observe that the mean of $ \vy | \vvt_{1} $ in \eqref{mod:mat-pol-obs} does not need any operation after the reordering. Moreover, instead of dealing with vectors and matrices of order $pT$, we simplify by working with $p$ vectors and matrices (also sparse) of order $T$ in \eqref{mod:mat-pol-ss1} and \eqref{mod:mat-pol-ss2}. 

The full conditional posterior distributions of the vectors $\vvt_k$, $k = 1, 2, \ldots, p$, are easier to handle than in the full vector $\vvt$. In order to simplify notations, we denote $ \mB = \mI - \mH$. Therefore,
\begin{equation}\label{fcond:vvtk-norm}
    \vvt_k | \fcond \sim N(\bar{\vmu}_k, \bar{\mP}_k^{-1}),
\end{equation}
where the precision matrix and the mean vector are, respectively, 
\[
\bar{\mP}_{k} = \left\{\begin{array}{ll}
\frac{1}{V} \mI + \frac{1}{W_1} \mH^{\T} \mH, & \text{ if } k = 1, \\
\frac{1}{W_{k-1}} \mB^{\T} \mB + \frac{1}{W_k} \mH^{\T} \mH, & \text{ if } k = 2, 3, \ldots, p,
\end{array}
\right.
\]
and
\[
\bar{\vmu}_k = \left\{\begin{array}{l}
\bar{\mP}_{1}^{-1} \co{ \dfrac{1}{V} \vy + \dfrac{\theta_{01}+\theta_{02}}{W_1} \ve_1 + \dfrac{1}{W_1} \mH^{\T} \mB \vvt_{2} }, \text{ if } k = 1, \\ \\
\bar{\mP}_{k}^{-1} \left[ \dfrac{1}{W_{k-1}} \mB^{\T} \mH \vvt_{k-1} + \dfrac{\theta_{0k}+\theta_{0(k+1)}}{W_k} \ve_{1} + \dfrac{1}{W_k} \mH^{\T} \mB \vvt_{k+1} \right], \text{ if } k = 2, \ldots, p-1, \\ \\
\bar{\mP}_{p}^{-1} \co{ \dfrac{1}{W_{p-1}} \mB^{\T} \mH \vvt_{p-1} + \dfrac{\theta_{0p}}{W_p} \ve_{1} }, \text{ if } k = p.
\end{array}
\right.
\]

Although it looks complicated, the sequential structure is simple to implement and can be easily generalized to the case where the data are not normally distributed, more efficiently than in model \eqref{mod:mat-genln-obs}-\eqref{mod:mat-gendln-ss}.

It remains to discuss the initial values $ \theta_{0k} $ and the variances $ V $ and $ W_k $, $ k = 1, 2, \ldots, p $. We assume independent priors. With respect to the initial values, if we consider the priors $ \theta_{0k} \sim N(\mu_{\theta_{0k}}, \sigma_{\theta_{0k}}^2) $, it is easy to see that the full conditional posterior is
\begin{equation}\label{fcond:theta0k}
\theta_{0k}|\fcond \sim N(\bar{\mu}_{0k}, \bar{\sigma}_{0k}^2),
\end{equation}
where
\[
\bar{\sigma}_{0k}^2 = \begin{cases}
\pa{\dfrac{1}{\sigma_{01}^2} + \dfrac{1}{W_1}}^{-1}, &  \text{ if } k = 1, \\
\pa{\dfrac{1}{\sigma_{0k}^2} + \dfrac{1}{W_{k-1}} + \dfrac{1}{W_{k}}}^{-1}, &  \text{ if } k = 2, 3, \ldots, p,
\end{cases}
\]
and
\[
\bar{\mu}_{0k} = \begin{cases}
\bar{\sigma}_{01}^2 \pa{\dfrac{\mu_{01}}{\sigma_{01}^2} + \dfrac{\theta_{11} - \theta_{02}}{W_1}}, & \text{ if } k = 1; \\ \\
\bar{\sigma}_{0k}^2 \pa{\dfrac{\mu_{0k}}{\sigma_{0k}^2} + \dfrac{\theta_{1(k-1)} - \theta_{0(k-1)}}{W_{k-1}} + \dfrac{\theta_{1k} - \theta_{0(k+1)}}{W_{k}}}, &  \text{ if } k = 2, 3, \ldots, p-1; \\ \\
\bar{\sigma}_{0p}^2 \pa{\dfrac{\mu_{0p}}{\sigma_{0p}^2} + \dfrac{\theta_{1(p-1)} - \theta_{0(p-1)}}{W_{k-1}} + \dfrac{\theta_{1p}}{W_{p}}}, &  \text{ if } k = p.
\end{cases}
\]

Instead of working with variances, we consider the precisions $ 1/V $ and $ 1/W_k $. Thus, assume that $ W_k^{-1} \sim \Gamma(\nu_{0k}, \eta_{0k}) $. One can see that the full conditional posterior is
\begin{equation}\label{fcond:wk}
W_k^{-1} | \fcond \sim \Gamma(\bar{\nu}_{0k}, \bar{\eta}_{0k}),
\end{equation}
with parameters
\begin{eqnarray*}
\bar{\nu}_{0k} & = & \nu_{0k} + \dfrac{T}{2}, \\
\bar{\eta}_{0k} & = & \eta_{0k} + \dfrac{1}{2} (\vvt_{k} - \vmu_k)^{\T}\mH^{\T} \mH (\vvt_{k} - \vmu_k),
\end{eqnarray*}
where $ \vmu_k $ is the same in \eqref{mod:mat-pol-ss1}-\eqref{mod:mat-pol-ss2} for $ k = 1, 2, \ldots, p $.

Similarly, if $ V^{-1} \sim \Gamma(\nu_{y}, \eta_{y}) $, then
\begin{equation}\label{fcond:v}
V^{-1}| \fcond \sim \Gamma(\bar{\nu}_{y}, \bar{\eta}_{y}),
\end{equation}
where
\begin{eqnarray*}
\bar{\nu}_{y} & = & \nu_{y} + \dfrac{T}{2}, \\
\bar{\eta}_{y} & = & \eta_{y} + \dfrac{1}{2} \sum_{t=1}^{T}(y_t - \theta_{t1})^2.
\end{eqnarray*}

All the posteriors above are conjugated, which allows the use of the Gibbs algorithm. In the next section, we deal with the case where the relationship between observation and state equations is nonlinear, which can be used to deal with the dynamic mixture weights (see Section \ref{sec:dynmixw}). This situation demands more computational efforts and our approach tends to simplify the development of the method.

The derivation of the results presented in this subsection is available in a supplementary material.

\subsection{Bayesian inference in nonlinear dynamic models}\label{sec:ndm}

In the case where the data are not normally distributed or the relationship between observation and state equations is nonlinear, Chan and collaborators proposed extensions to the precision-based algorithm \citep[see, e.g.,][]{Chan-Koop-Potter-2013}. The idea is to apply the accept-reject Metropolis-Hastings (ARMH) algorithm \cite[see][]{ChibGreenberg1995} to the whole vector of states. This method is described in more details in \cite{Chan-Strachan-2012}.

Another benefit of considering the reordering of the vector of states, as proposed in Section \ref{sec:idgp}, is the ability to deal with nonlinear dynamic models. The main reason is that extensions like those cited above tend to be more computationally intensive. Therefore, the smaller the vector of states is, the better. Thus, instead of dealing with the whole vector of states as in the approach of Chan and collaborators, in our proposal one can deal with only the first component of the state vector, $ \vvt_{1} $. The remaining vectors $ \vvt_{k} $, $ k = 2, 3, \ldots, p $, will have full conditional posteriors as in \eqref{fcond:vvtk-norm}, which is more easily calculated.

One problem of considering the ARMH algorithm as presented in \cite{Chan-Strachan-2012} and \cite{Chan-Koop-Potter-2013} is that, depending on the complexity (e.g., the distribution of the observed data and/or the size $ T $ of the series), the algorithm tends to have difficulties in providing a good acceptance rate, which might interfere in the performance of the MCMC. In a few numerical experiments (not reported here), involving ``simple cases'' of Bernoulli data, the algorithm was problematic. Because of this issue, and exploiting an interesting property of the joint (prior) distribution of $ \vvt_1 $, we consider two alternatives in this work: (i) the component-wise Metropolis-Hastings (CWMH) algorithm (MH algorithm for each element of $ \vvt_{1} $); and (ii)  for the specific case of Bernoulli data, the use of the probit link function. The latter case is not as general as the former, but it is efficient when the data in hand is Bernoulli (and will be useful to estimate the dynamic mixture weights in next section).

In a general situation, let the likelihood function be written as $ p(\vy | \va) = \prod_{t=1}^{n} p(y_t | \alpha_t) $. Also, assume that there exists a (link) function $ \TT: A \longrightarrow \R $, which is continuous and bijective, where $ A $ corresponds to the parameter space of the $ \alpha_{t} $'s. Thus, the spate parameters $ \vvt $ are related to $ \va $ by $ \TT(\alpha_t) = \theta_{t1} $, $ t = 1, 2, \ldots, T $. Therefore, once the full conditional posterior of $ \vvt_{1} $ is drawn, one can have $ \alpha_t = \TT^{-1}(\theta_{t1}) $, where $ \TT^{-1} $ denotes the inverse function of $ \TT $.

\subsubsection{Component-wise Metropolis-Hastings}\label{sec:cwmh}

The idea of CWMH might look computationally intensive. However, when dealing with polynomial DLMs, several calculations are simplified. An interesting property that can be used, closely related to results in \cite{Rue-Held-2005}, is presented below.

\begin{theorem}\label{theo:fullcond-theta1}
Let $ \vvt_1 $ be defined as in \eqref{thet:markov} and denote $ \vvt_{(t)1} = (\theta_{11}, \ldots, \theta_{t-1,1}, \theta_{t+1,1}, \ldots, \theta_{T1})^{\T} $. Thus
\[
\theta_{t1} | \vvt_{(t)1}, \vvt_{2} \sim N(\mu_{t1}^{*}, \tau_{t}^2),
\]
where
\[
\mu_{t1}^{*} = \begin{cases}
\dfrac{1}{2}\co{\pa{\theta_{t+1,1} - \theta_{t2}} + \pa{\theta_{t-1,1} + \theta_{t-1,2}}}, & \text{if } t = 1, \ldots, T - 1, \\
\theta_{T-1,1} + \theta_{T-1,2}, & \text{if } t = T,
\end{cases}
\]
and
\[
\tau_{t}^2 = \begin{cases}
\dfrac{W_1}{2}, & \text{if } t = 1, \ldots, T - 1, \\
W_1, & \text{if } t = T.
\end{cases}
\]
\end{theorem}

By this theorem, it is clear that the conditional distribution of $ \theta_{t1} | \vvt_{(t)1}, \vvt_{2} $ demands only $ O(T) $ operations, showing how simplified the process can be. Furthermore, observe that $ \mu_{t1}^{*} $ above can be seen as a prediction for $ \theta_{t1} $. For $ t = 1, \ldots, T - 1 $, $ \mu_{t1}^{*} $ is an average of conditional expectations of $ \theta_{t1} $ in the $ t $-th and $ (t+1) $-th equations in the local level, whereas $ \mu_{T1}^{*} $ can be interpreted as a conditional expectation of $ \theta_{T1} $ in the $ T $-th equation. The proof of the theorem is available in a supplementary material.

Thus, based on Theorem \ref{theo:fullcond-theta1}, one can derive the full conditional posterior distribution as
\[
p(\theta_{t1} | \fcond) \propto \varphi(\theta_{t1} | \mu_{t1}^{*}, \tau_{t}^2) p(y_t | \TT^{-1}(\theta_{t1})),
\]
where $ t = 1, 2, \ldots, T $ and $ \varphi(x | a, b) $ denotes the probability density function of a normal \rv\ with mean $ a $ and variance $ b $, evaluated at $ x $. Observe by the posterior above that the CWMH is quite general and fits to any nonlinear case, with any (bijective) link function $ \TT $.

As a proposed distribution, we consider a random walk, with variance adapted during the MCMC iterations as in \cite{Roberts-Rosenthal-2009}. Then, for each $ t = 1, 2, \ldots, T $, one can draw a candidate $ \theta_{t1}^* \sim N(\theta_{t1}, \varsigma_t^2) $, which will or will not be accepted in a Metropolis-Hastings procedure. The variance $ \varsigma_t^2 $ can be increased/reduced in order to ensure an acceptance rate of 0.44. Basically, during the MCMC, after the $ n $-th ``batch'' of 50 iterations, the authors suggest adding or subtracting the standard deviation in log-scale by $ \min(0.01, n^{-1/2}) $. For more details, see the aforementioned reference.

\subsubsection{The probit link for Bernoulli data}\label{sec:probit}

For the specific case of Bernoulli data, one can also consider another efficient way of sampling the $ \alpha_{t} $'s almost jointly. One can consider the probit link function to apply the proposal of \cite{Albert-Chib-1993}. 

Here we use $ \TT \equiv \Phi^{-1} $, the inverse of the cumulative distribution function of a standard normal \rv\ Basically, we have that $ \alpha_t = \Phi(\theta_{t1}) $, $ t = 1, 2, \ldots, T $. The trick consists of writing $ \alpha_t $ in a GLM context, where the ``design matrix'' is an identity of order $ T $, denoted by $ \mI $. 

In this case, there exist $ T $ latent \rv's $ v_1, v_2, \ldots, v_T $, where the $ v_t $'s are independent, with $ v_t \sim N(\theta_{t1}, 1) $, such that $ y_t = 1 $ if $ v_t > 0 $, and $ y_t = 0 $ otherwise. Thus, one can see that $ \alpha_t = \p(y_t = 1) = \Phi(\theta_{t1}) $. Therefore, based on \eqref{mod:mat-pol-ss1}, it is easy to see that
\begin{eqnarray*}
v_t | y_t = 0, \theta_{t1} & \sim & N(\theta_{t1}, 1) \text{ truncated at the right by } 0, \\
v_t | y_t = 1, \theta_{t1} & \sim & N(\theta_{t1}, 1) \text{ truncated at the left by } 0.
\end{eqnarray*}
The full conditional posterior for $ \vvt_{1} $ is
\begin{equation}\label{fcond:theta1-vector-probit}
\vvt_{1} | \vv, \fcond \sim N(\bar{\vmu}_{1}, \bar{\mP}_{1}^{-1}),
\end{equation}
where
\begin{eqnarray*}
\bar{\mP}_{1} & = & \mI + \frac{1}{W_1} \mH^{\T} \mH, \\
\bar{\vmu}_{1} & = & \bar{\mP}_{1}^{-1} \co{ \vv + \dfrac{\theta_{01}+\theta_{02}}{W_1} \ve_1 + \dfrac{1}{W_1} \mH^{\T} \mB \vvt_{2} },
\end{eqnarray*}
with $ \vv = (v_1, v_2, \ldots, v_T)^{\T} $. Compare the similarity of \eqref{fcond:theta1-vector-probit} with \eqref{fcond:vvtk-norm}. Basically, the full conditional posterior of $ \vvt_{1} $ here has one more step, which corresponds to the generation of the latent variables $ \vv $. The consequence is an algorithm almost as efficient as in the case of the dynamic Gaussian linear model.

\section{The Gaussian mixture model}\label{sec:mixmodel}

In this paper, we examine a dynamic Bayesian mixture of independent Gaussian distributions, with unknown means and precisions. The model can be specified as 
\begin{equation}\label{mod:dgmm}
\begin{aligned}
y_t | z_t, \vmu, \vphi &\sim N[z'_t {\vmu}, (z'_t {\vphi})^{-1}], \\
z_t | \valpha_t &\sim Cat(\valpha_t),
\end{aligned}
\end{equation}
$ t = 1, 2, \ldots, T $, where the $y_t$'s are observed components, and the $z_t$'s are latent components, that indicate the normal population to which the $ t $-th observation belongs. In other words, $ z_t = (z_{1t}, \ldots, z_{Kt})' $ is a vector such that $ z_{kt} = 1 $, if $ y_t $ belongs to the $ k $-th normal population, and zero otherwise. The vector $ \valpha_t = (\alpha_{1t}, \ldots, \alpha_{Kt})' $ corresponds to the dynamic mixture weights, that is able to vary along time $ t $, with $ \alpha_{kt} \geq 0 $ and $ \sum_{k=1}^{K} \alpha_{kt} = 1 $. Each weight $ \alpha_{kt} $ characterizes the probability that $ z_{kt} = 1 $. Moreover, $ \vmu = (\mu_1, \ldots, \mu_K)' $ and $ \vphi = (\phi_1, \ldots, \phi_K)' $ represent the mean and precision vectors, respectively. Also, $Cat(\cdot)$ can be used to denote a categorical variable.

The main goal in this section is  the estimation of $ \vmu $, $ \vphi $, the $ z_t $'s and the $ \valpha_t $'s. In order to derive the posterior distribution of these parameters, we assume that $ p(\vy | \vz, \vmu, \vphi) = \prod_{t=1}^T p(y_t | z_{t}, \vmu, \vphi) $ and $ p(\vz | \valpha_1, \ldots, \valpha_T) = \prod_{t=1}^T p(z_t | \valpha_t) $.  
This means that $ y_t $ is a time series and its dependence structure is mostly related to the functional behavior of the probability of the mixture sequence $ \valpha_t $.

The case where $ \valpha_t \equiv \valpha $ corresponds to the ordinary Gaussian mixture model, and it is taken into account in the next subsection (which does not interfere in the results for the component parameters). There is a vast literature for this setup. A comprehensive survey involving finite mixture models, under several scenarios, is presented in \cite{FruhwirthSchnatter2006}.

It is a usual practice to postulate independent prior distributions for the component parameters $ \vmu $ and $ \vphi $, i.e., $ p(\vmu, \vphi) = \prod_{k=1}^Kp(\mu_k) p(\phi_{k}) $. Examples of works employing independent priors are \cite{EscobarWest1995} and \cite{RichardsonGreen1997}. We consider in this paper the priors $ \mu_k \sim N(\mu_{0k}, \sigma_{0k}^2) $ and $ \phi_{k} \sim \Gamma(\nu_{0k}, \eta_{0k}) $, $ k = 1, \ldots, K $. When $ \valpha_t \equiv \valpha $, the prior of the mixture weights vector $ \valpha $ is usually a Dirichlet process, $ \valpha \sim Dir(\vect{e}_0) $, which is assumed to be independent of $ \vmu $ and $ \vphi $.

\subsection{Full conditional posterior distributions of the component parameters}\label{sec:fcond-comp}
In order to get the full conditional distributions, we begin by specifying the joint distribution of the observations, latent quantities and parameters:
\[ p( {\vy}, {\vz},  {\vmu}, {\vphi},  \valpha)= p(\vy| \vmu, \vphi, \vz) p(\vmu,\vphi) p(\vz|\valpha) p(\valpha). \]

We denote by $\fcond$ the set of all remaining variables to be considered for the posterior in use. It is straightforward to obtain that:
\begin{enumerate}
\item[(i)] the conditional posterior distribution for each mean and precision value are, respectively,
\begin{eqnarray}
\mu_k | \vy, \fcond & \sim & N(\bar{\mu}_{k}, \bar{\sigma}_{k}^2), \label{fcond:muk} \\
\phi_k | \vy, \fcond & \sim & \Gamma(\bar{\nu}_{k}, \bar{\eta}_{k}), \label{fcond:preck} 
\end{eqnarray}
where
\[
\begin{array}{rclllrcl}
\bar{\sigma}_{k}^2 & = & \pa{T_k \phi_k + 1/\sigma_{0k}^2}^{-1}, &&& \bar{\nu}_{k} & = & \nu_{0k} + T_k/2, \\
\bar{\mu}_{k} & = & \bar{\sigma}_{k}^2\pa{s_k \phi_k + \mu_{0k}/\sigma_{0k}^2}, &&& \bar{\eta}_{k} & = & \eta_{0k} + v_k,
\end{array}
\]
with $ T_k = \#\{z_{kt} = 1, t = 1, 2, \ldots, T\} $, $ s_k = \sum_{\{t:z_{kt}=1\}} y_t $ and $ v_k = \sum_{\{t:z_{kt}=1\}} (y_t - \mu_k)^2 $, $ k = 1, \ldots, K $;

\item[(ii)] the conditional posterior distribution of the latent categorical variable is\linebreak
$p(z_{kt}=1|\vy, \fcond) \propto \alpha_k \varphi(y_t|\mu_k, \phi_k^{-1}) $,
where $\varphi(x|a,b)$ denotes the probability density function of a normal \rv\ with mean $a$ and variance $b$.
Then, it follows that
\[
P(z_{kt} = 1 | \vy, \fcond) = \frac{\alpha_{k} \varphi(y_t | \mu_k,\phi_{k}^{-1})} {\Sigma_{k=1}^K \alpha_{k} \varphi(y_t | \mu_k, \phi_{k}^{-1})},
\]
$ k=1, \ldots, K; $
\item[(iii)] for the sake of information, the conditional posterior of the mixture weights is $ \valpha | \vy, \fcond \sim Dir(\vect{e}_1) $, where $ \vect{e}_1 = \vect{e}_0 + (T_1, \ldots, T_K)' $.
\end{enumerate}

A frequent issue involving mixture problems is label switching. There are several studies suggesting solutions to this kind of problem \citep[more details in][]{FruhwirthSchnatter2006}. Here, we consider a simple solution: the pairs $ (\mu_k, \phi_k) $ are ordered under the constraint\linebreak $ \mu_k < \mu_{k+1} $.

As mentioned before, the full conditional posteriors in the case where the weights are dynamic are the same as in (i) and (ii) above. Thus, it remains to study situations where the mixture weights vary over time. 

\subsection{Bayesian estimation of the dynamic mixture weights}\label{sec:dynmixw}
For the sake of simplicity, we consider the case where $ K = 2 $, i.e., a dynamic Gaussian mixture of two groups. In this scenario, $ z_t $ is equivalent to a Bernoulli \rv\ with parameter $ \alpha_t $, the dynamic mixture weight. Therefore, we focus on the general case where $\alpha_t$ varies throughout time. Thus, the Gaussian mixture model in \eqref{mod:dgmm} can be rewritten as
\begin{eqnarray*}
y_t | z_t, \vmu, \vphi & \sim & N(m^*, s^*), \\
z_t | \alpha_t &\sim & \bern(\alpha_t), \quad t = 1, 2, \ldots, T,
\end{eqnarray*}
where $ m^* = z_t\mu_1 + (1 - z_t) \mu_2 $ and $ s^* = z_t \phi_1^{-1} + (1 - z_t) \phi_2^{-1} $. Moreover, we assume that the dynamic evolution of the $ \alpha_t $'s behaves according to a nonlinear dynamic model, as discussed in Section \ref{sec:ndm}.

The component parameters can be easily estimated according to \eqref{fcond:muk} and \eqref{fcond:preck}. Also, based on item (ii) in Section \ref{sec:fcond-comp}, it is easy to generalize and see that the full conditional posterior of $ z_t $ can be written as
\begin{equation}\label{fcond:zt}
\begin{aligned}
&z_t|y_t, \fcond \sim \bern\pa{\alpha_t^{*}}, \\
&\alpha_t^*  = \dfrac{\alpha_t \varphi(y_t | \mu_2, \phi_2^{-1})}{(1 - \alpha_t) \varphi(y_t | \mu_1, \phi_1^{-1}) + \alpha_t \varphi(y_t | \mu_2, \phi_2^{-1})},
\end{aligned}
\end{equation}
$ t = 1, 2, \ldots, T $. Therefore, it only remains to deal with the dynamic mixture weights.

Once the latent categorical variables $ z_t $'s are generated, one can proceed to estimate the $ \alpha_t $'s as in the nonlinear dynamic model, with a Bernoulli response. Therefore, the full conditional posterior of the dynamic mixture weights can be derived according to the procedures described in Sections \ref{sec:cwmh} and \ref{sec:probit}. In the former case, a natural candidate as link function is the logit, where $ \logit(x) = \log[x/(1-x)] $, for $ 0 < x < 1 $.

\subsubsection{Gibbs sampling algorithm}\label{sec:gibbs}
Once we have in hand the full conditional posterior distributions, we can generate the MCMC for the problem. Thus, posterior draws can be obtained by sequentially sampling as below:

\begin{enumerate}
\item Generate the means and precisions of the mixture parameters $ (\mu_k, \phi_{k}) $, $ k = 1, 2 $, as in \eqref{fcond:muk} and \eqref{fcond:preck}. After generation, order the pairs under the constraint\linebreak $ \mu_1 < \mu_2 $;
\item Generate an independent sample of $ z_t $, $ t = 1, 2, \ldots, T $, as in \eqref{fcond:zt};
\item For $ k $ from $ p $ to $ 2 $:
\begin{enumerate}
\item Generate $ \theta_{0k}  $ as in \eqref{fcond:theta0k};
\item Generate $ W_k $ as in \eqref{fcond:wk};
\item Generate $ \vvt_{k} $ as in \eqref{fcond:vvtk-norm};
\end{enumerate}
\item Generate $ \theta_{01} $ as in \eqref{fcond:theta0k};
\item Generate $ W_1 $ as in \eqref{fcond:wk};
\item Generate $ \vvt_{1} $ as in Section \ref{sec:cwmh} for the logit link, or Section \ref{sec:probit} for the probit link (using the categorical $ z_t $'s instead of $ y_t $'s in the refereed sections);
\item Calculate $ \alpha_t = \TT^{-1}(\theta_{t1}) $, $ t = 1, 2, \ldots, T $.
\end{enumerate}

\section{Artificial data}\label{sec:simul-data}

In this section we evaluate the performance of the method proposed in Section \ref{sec:dinmixmodel} using simulated data. Motivated by the arguments in Section \ref{sec:idgp} (second paragraph), second order dynamic polynomial models are considered for this task. In this case, the mixture weight evolve over time, following traditional patterns found in the literature.

We focus on the diversity of shapes, in order to see how the method performs under different scenarios. Therefore, we consider four different dynamic behaviors for $ \alpha_t $, which are presented here scaled in the unit interval:
\begin{enumerate}[(1)]
\item Linear behavior: $$ \alpha_t^{(1)} = 0.1 + 0.8t; $$
\item Parabolic behavior: $$ \alpha_t^{(2)} = 3(t - 0.5)^2 + 0.125; $$
\item Sinusoidal behavior: $$ \alpha_t^{(3)} = \cos(2\pi(t + \pi))/2.5 + 0.5; $$
\item Stepwise behavior: $$ \alpha_t^{(4)} = 0.2 \1_{[0,0.3)}(t) + 0.8 \1_{[0.3,0.7)}(t) + 0.3 \1_{[0.7,1)}(t), $$ where $ \1_A(t) $ is an indicator function, which is one, if $ t \in A $, and zero, otherwise.
\end{enumerate}

The initial information of the state equations, defined as $\theta_{01}$ and $\theta_{02}$, is described through independent standard normal distributions, which is enough to provide a relatively vague initial information regarding $ \alpha_1 $ (the dynamic mixture weight of instant one). In other words, after applying the transformation (logit or probit), one can have the initial probability of the $ \alpha_t $'s in a range close to the unit interval. 

Unlike \cite{Biller-Fahrmeir-1997}, we simplify the structure of the precision innovations by taking into account independent priors of the form
\begin{eqnarray*}
1/W_1 & \sim & \Gamma(0.01, 0.01), \\
1/W_2 & \sim & \Gamma(0.01, 0.01).
\end{eqnarray*}
These two priors will provide precision parameters with mean 1 and variance 100. 

The MCMC chains were developed with 220,000 iterations for each parameter. From these chains we discarded a burn-in of size 20,000 and took observations with a lag of size 200, resulting in a final chain of 1,000 values. The point estimates considered here are the medians (based on the absolute risk). 

\subsection{Mixture data}\label{sec:mixdata}

In this study we focus on the mixture of two normally distributed groups of the kind
\[
y_t = (1-z_t) x_{1t} + z_t x_{2t},
\]
where $ z_t|\alpha_t \sim \bern(\alpha_t) $, $ x_{1t}  \sim N(0, 0.25) $, $ x_{2t} \sim N(2, 0.25) $ (which means\linebreak $ \phi_1 = \phi_2 = 4 $). With respect to $ \alpha_t $, we consider the cases of $ \alpha_t^{(k)} $, $ t = 1, 2, \ldots, T $, $ k = 1, 2, 3, 4 $ described at the beginning of the section. We generated datasets of sizes $ T = 200, 400, 800 $. Since the results were similar, we present only the case where $ T = 400 $.

\begin{figure}
\centering
\includegraphics[angle=0,width=0.4\linewidth]{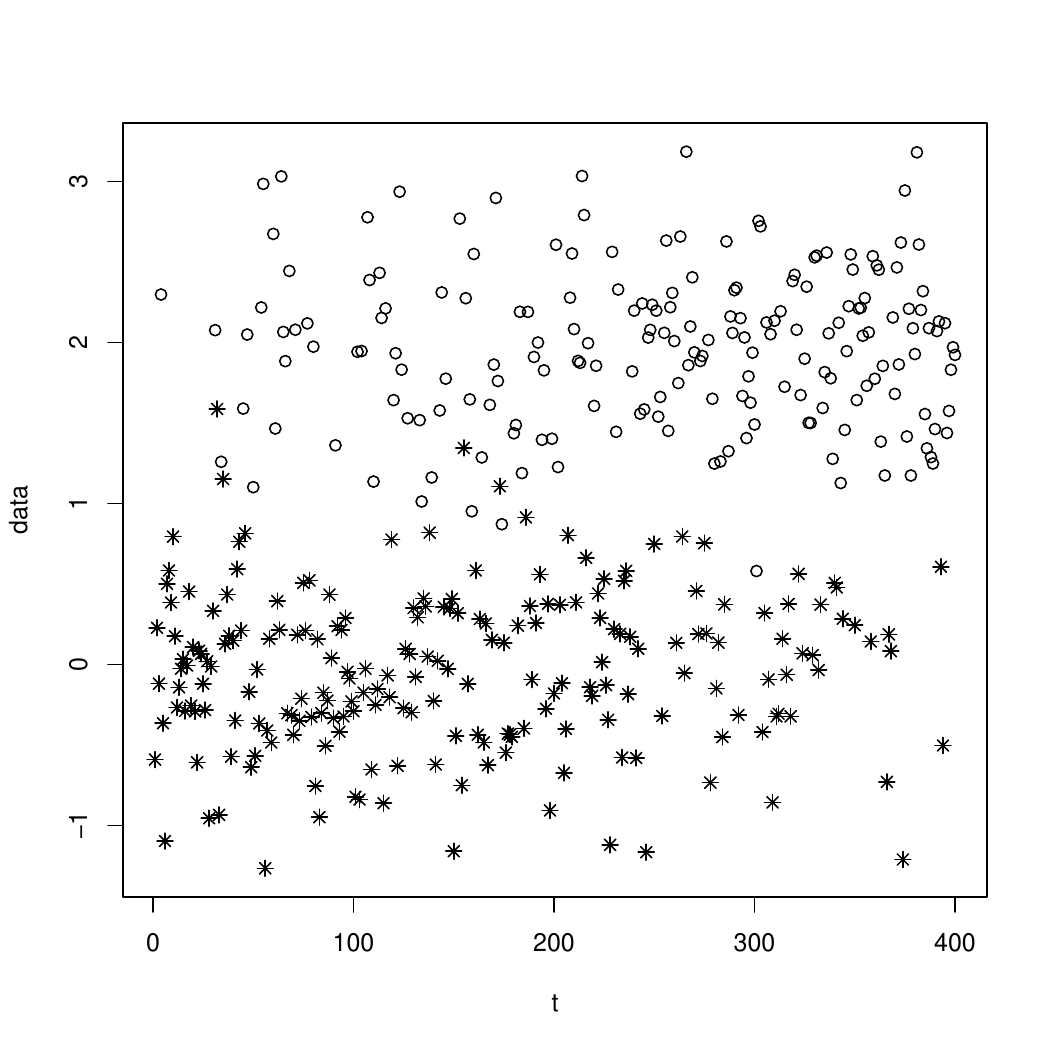}
\includegraphics[angle=0,width=0.4\linewidth]{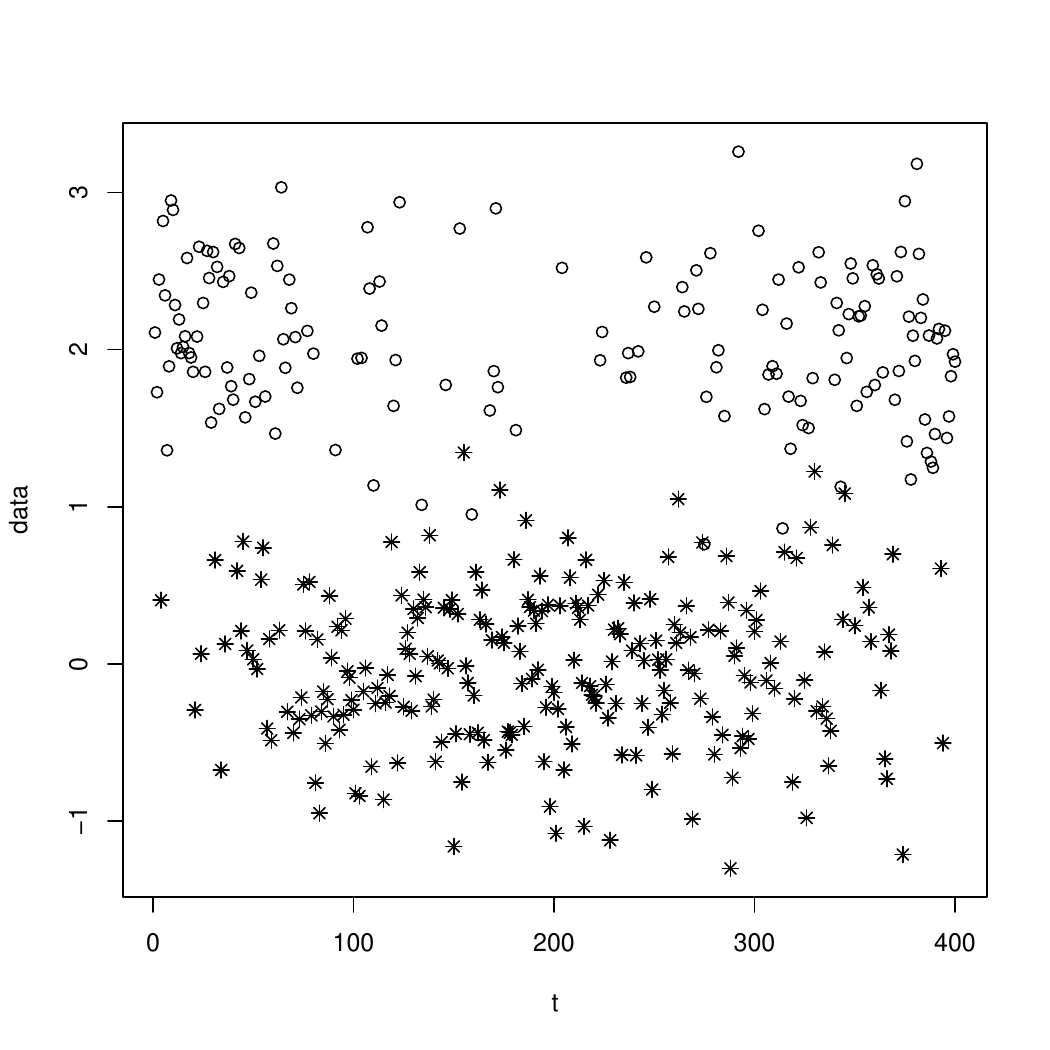} \\
\includegraphics[angle=0,width=0.4\linewidth]{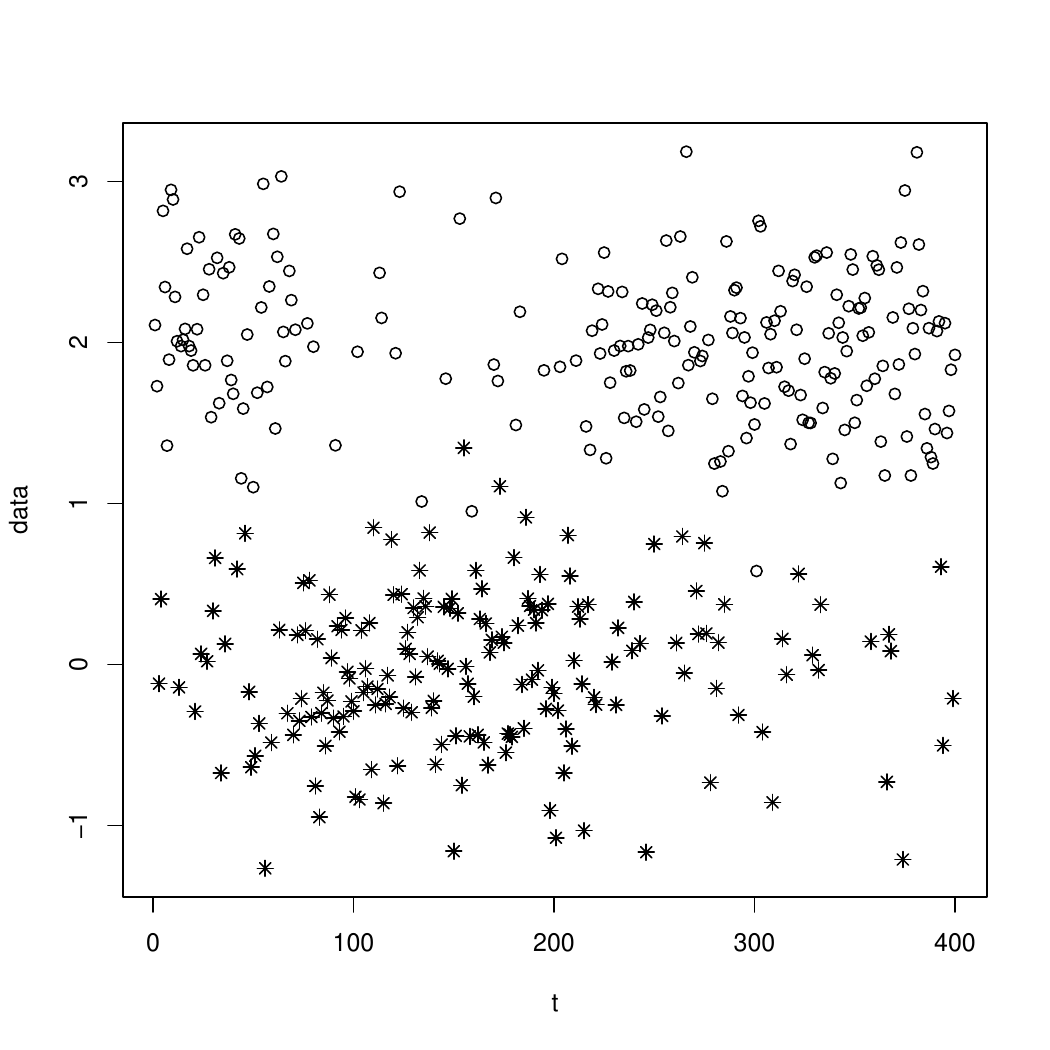}
\includegraphics[angle=0,width=0.4\linewidth]{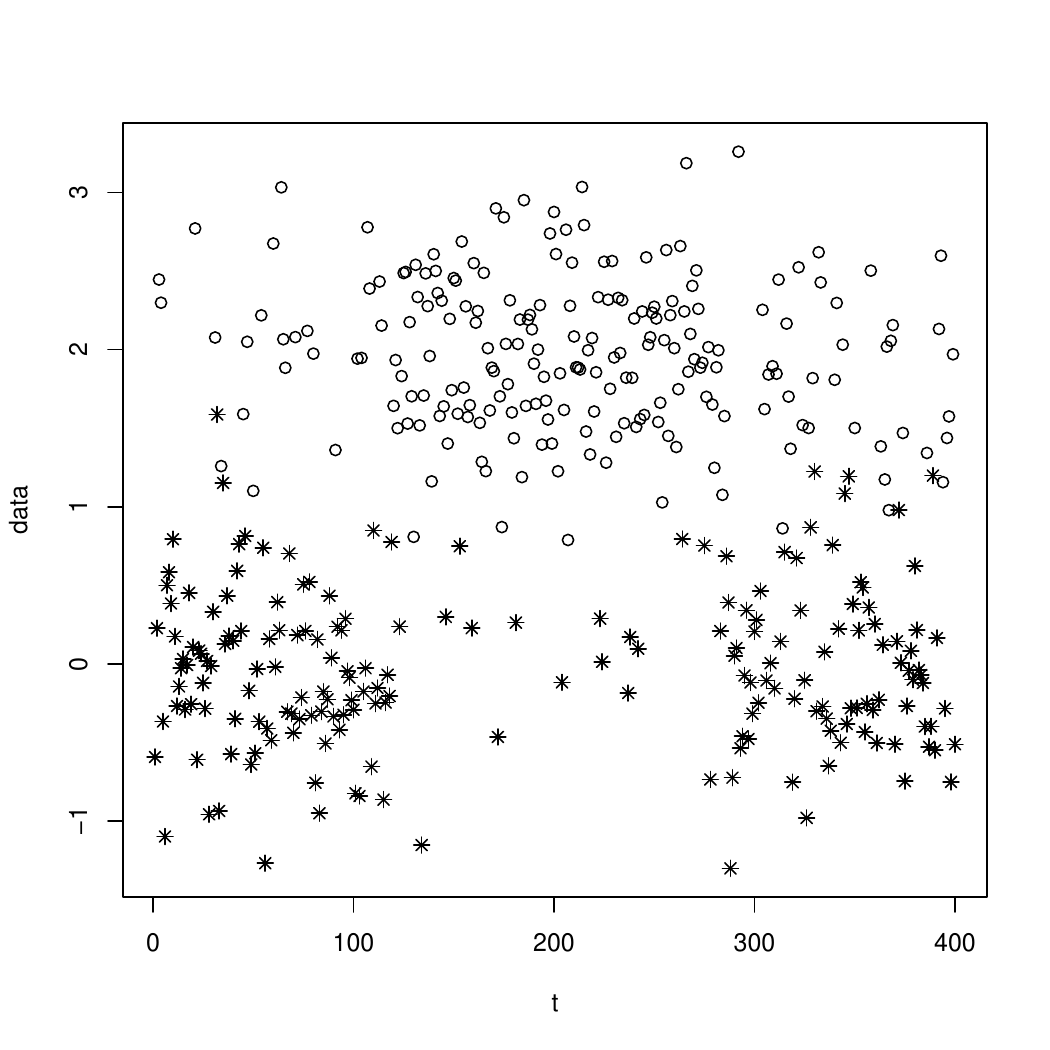}
\caption{Datasets of size $ T = 400 $ generated based on the Gaussian mixture, by considering mixture weights with linear behavior (top left), parabolic behavior (top right), sinusoidal behavior (bottom left) and steps behavior (bottom right). The symbol $ * $ represents $ y_t \stackrel{d}{=} x_{1t} $ and $ \circ $ means $ y_t \stackrel{d}{=} x_{2t} $.}
\label{fig:mixdatasets}
\end{figure}

The data generated are presented in Figure \ref{fig:mixdatasets}. Observe the complexity of identifying the real dynamic mixture weights, even with the groups being highlighted (which does not happen in practice).

With respect to the priors for the component parameters, we considered\linebreak $ \mu_{1} \sim N(q_1, 10s^2) $, $ \phi_1 \sim \Gamma(0.01, 0.01) $, $ \mu_{2} \sim N(q_3, 10s^2) $ and $ \phi_2 \sim \Gamma(0.01, 0.01) $, where $ q_1 $ and $ q_3 $ correspond to the 1st and 3rd quartile of the observed data, respectively, and $ s^2 $ represents the sample variance. We maintained the same prior applied for other precision parameters. Moreover, the priors for the means are relatively vague, with their means respecting the amplitude of the data. In this scenario, the proposed priors also took into account the constraint to avoid label switching (step 1 of Section \ref{sec:gibbs}).

For the generated mixture data, according to each mixture weight $ \alpha^{(1)} $ -- $ \alpha^{(4)} $, MCMC chains were run by the Gibbs algorithm indicated in Section \ref{sec:gibbs}. We considered both link functions, probit and logit, in the estimation of the $ \alpha_t $'s.

\begin{figure}
\centering
\includegraphics[angle=0,width=0.3\linewidth]{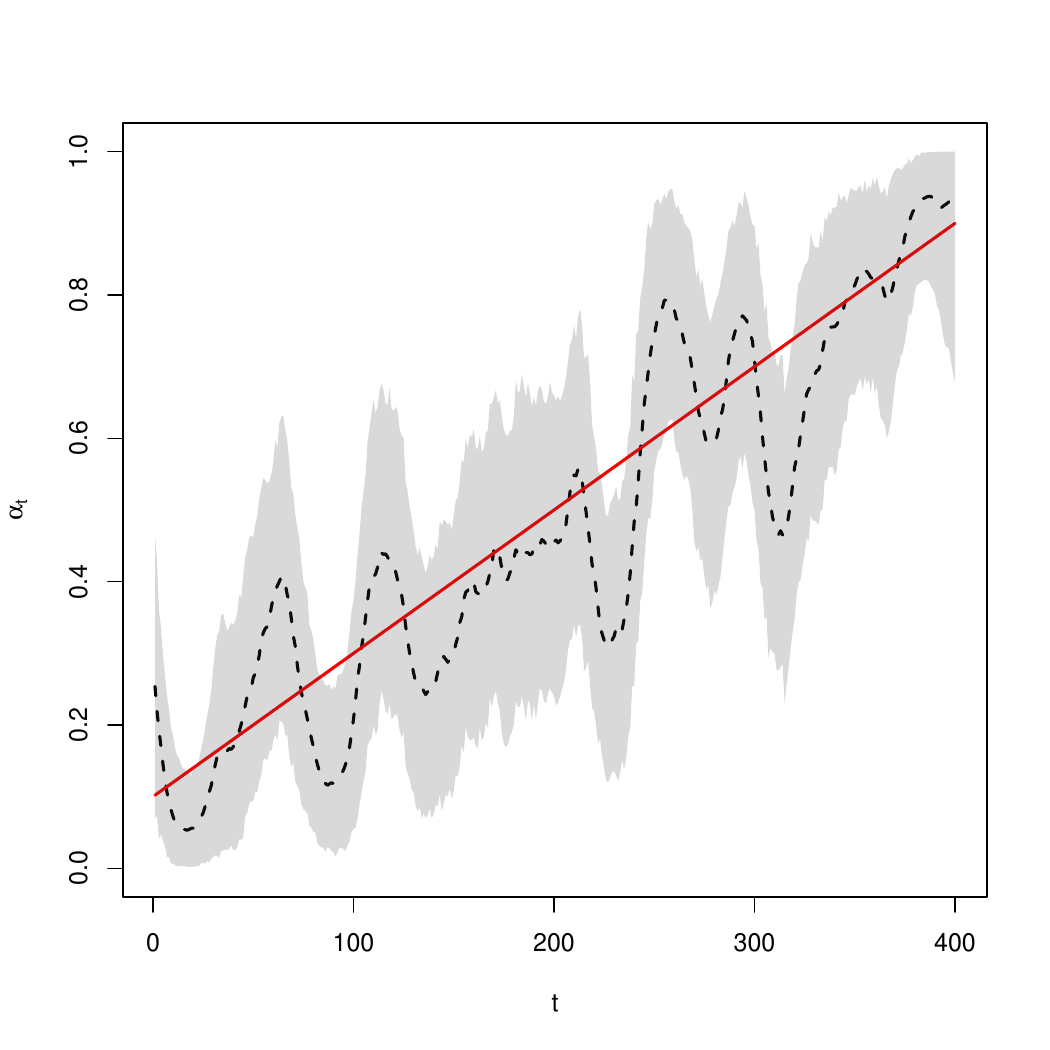}
\includegraphics[angle=0,width=0.3\linewidth]{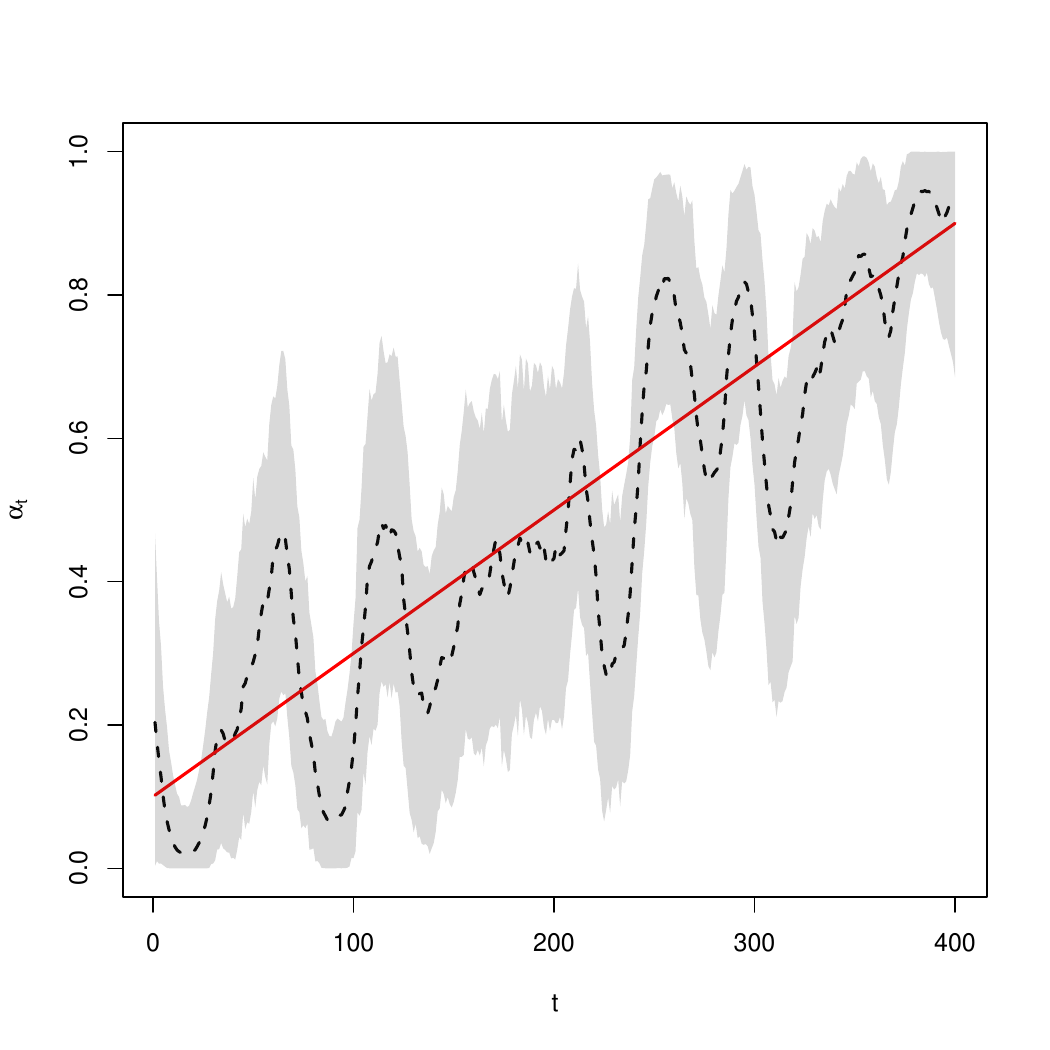} \\
\includegraphics[angle=0,width=0.3\linewidth]{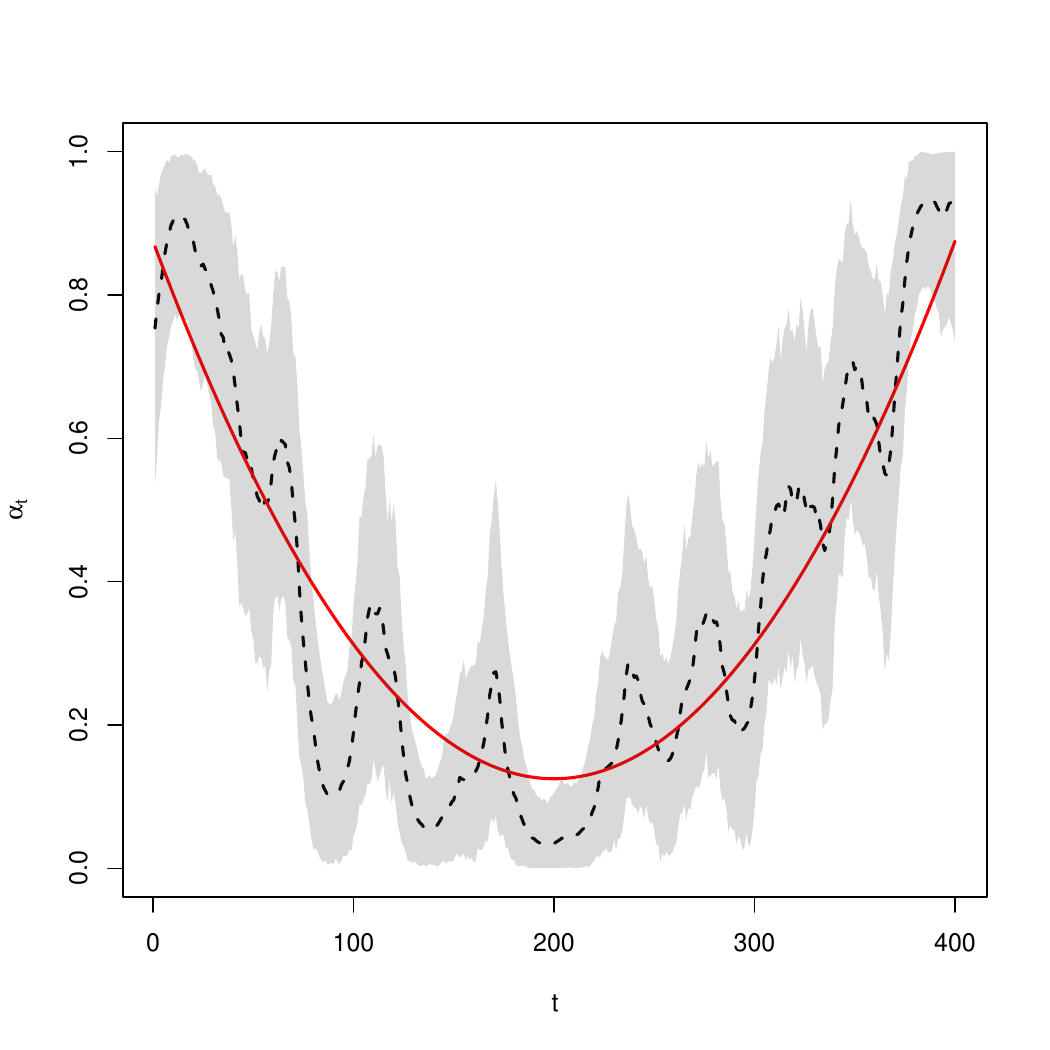}
\includegraphics[angle=0,width=0.3\linewidth]{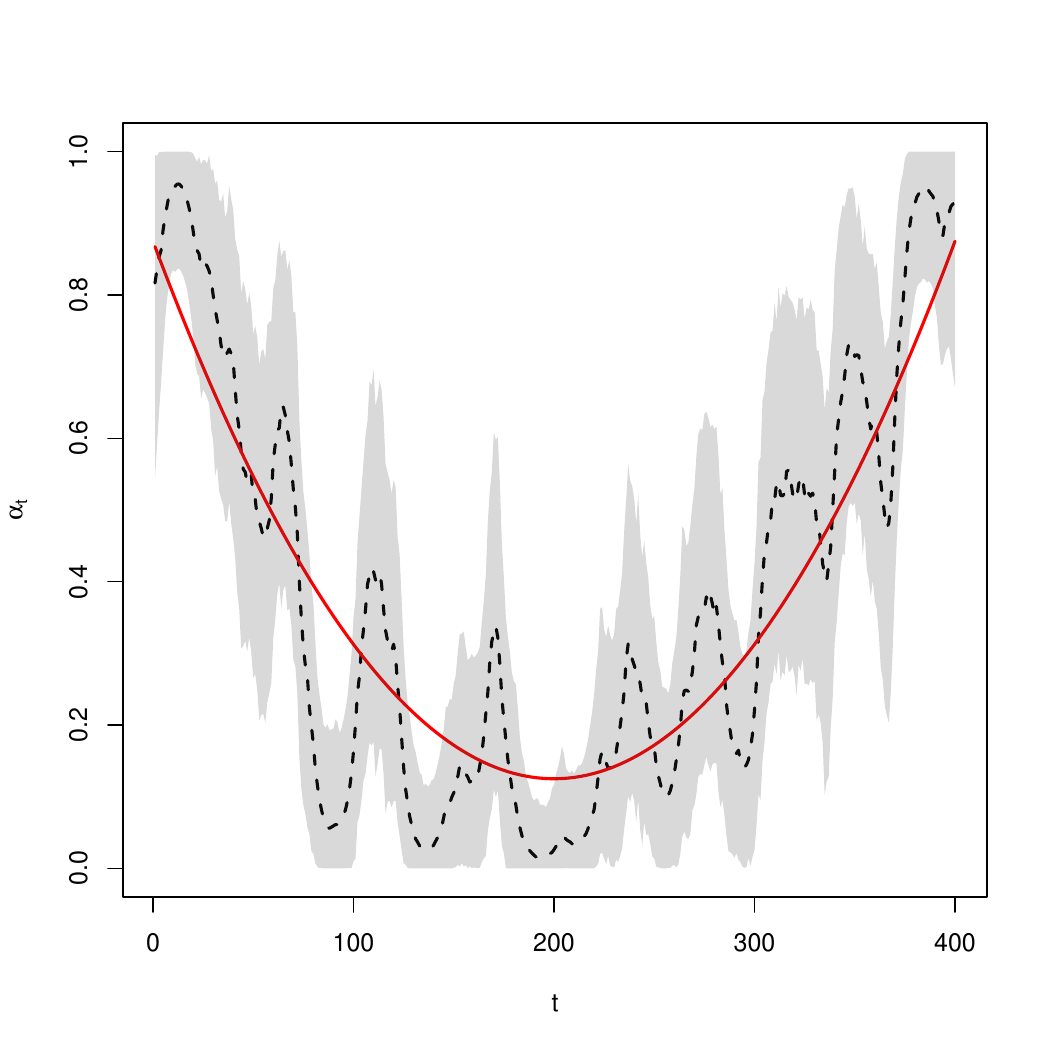} \\
\includegraphics[angle=0,width=0.3\linewidth]{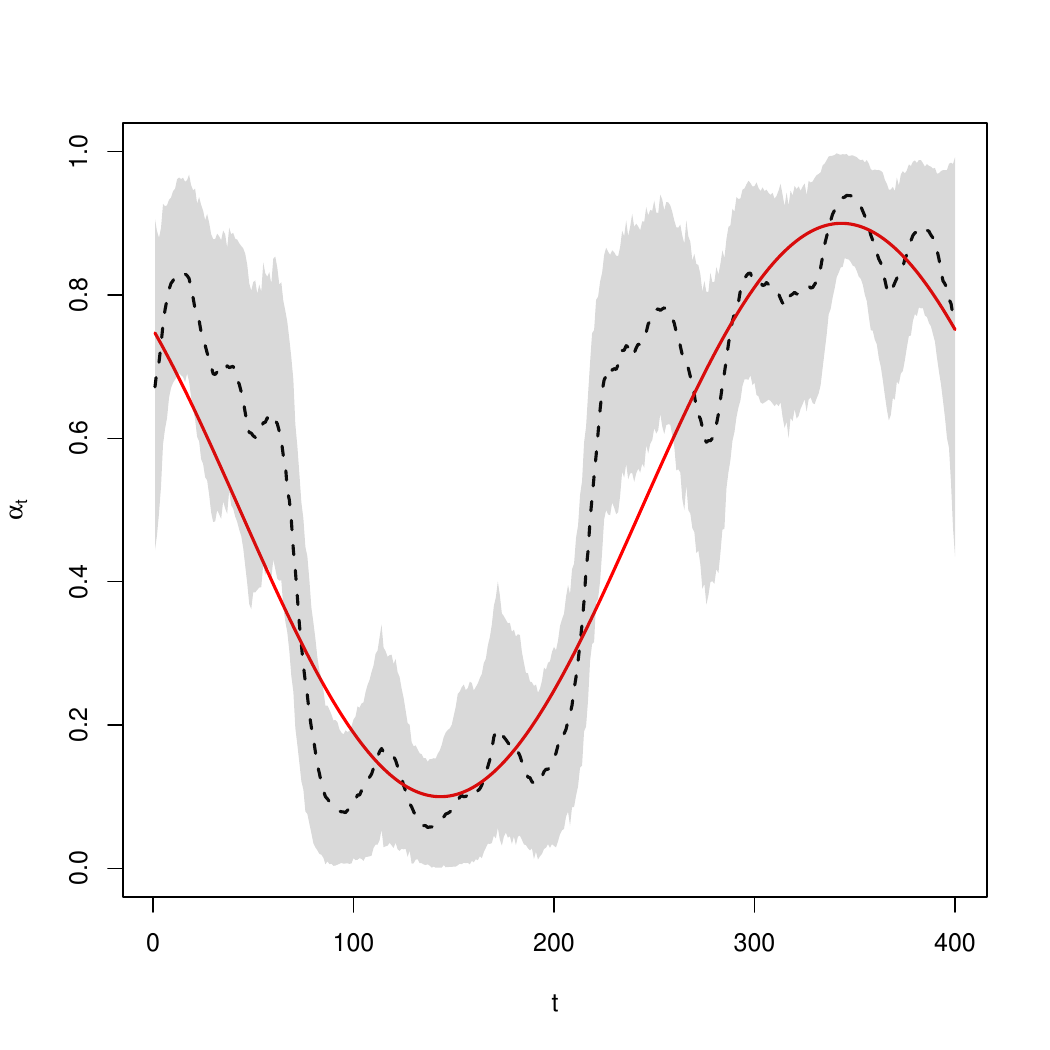}
\includegraphics[angle=0,width=0.3\linewidth]{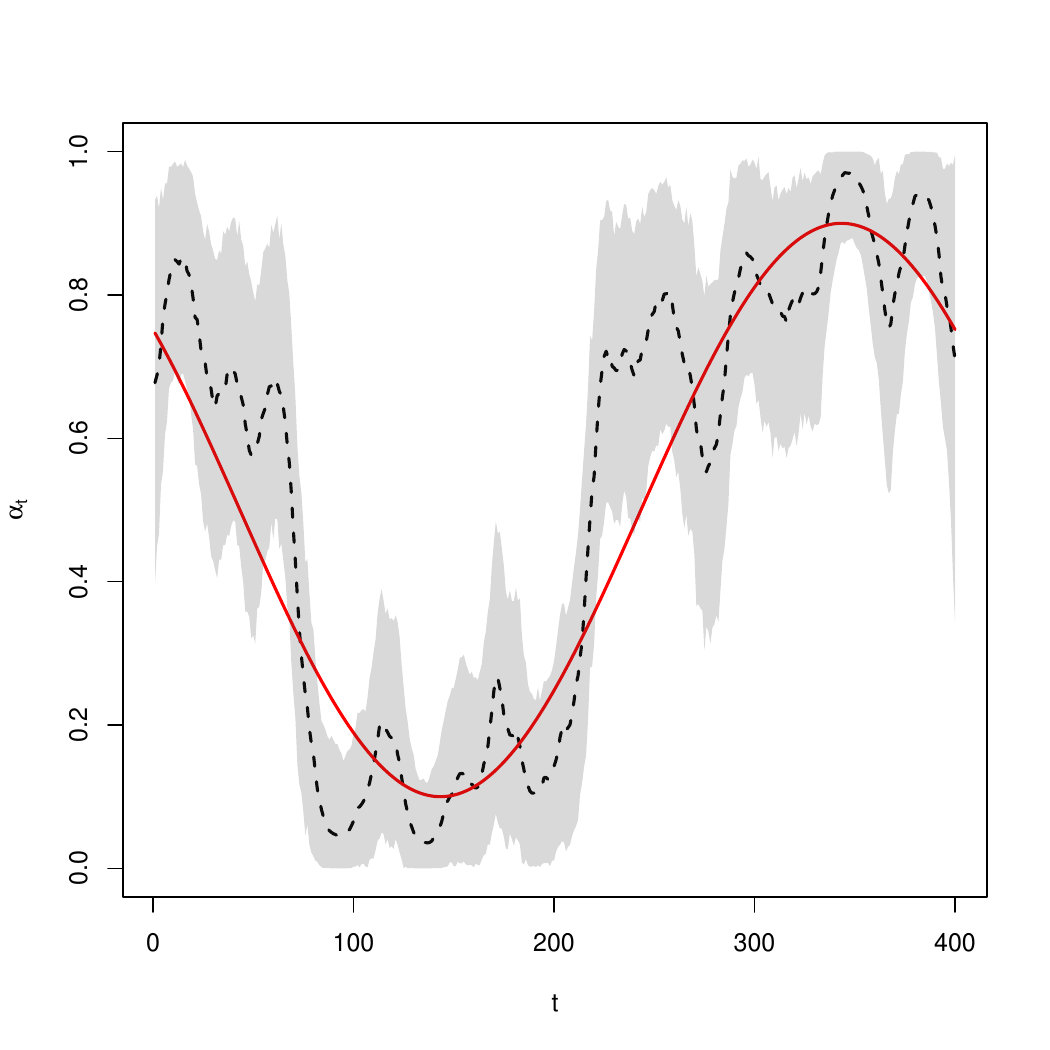} \\
\includegraphics[angle=0,width=0.3\linewidth]{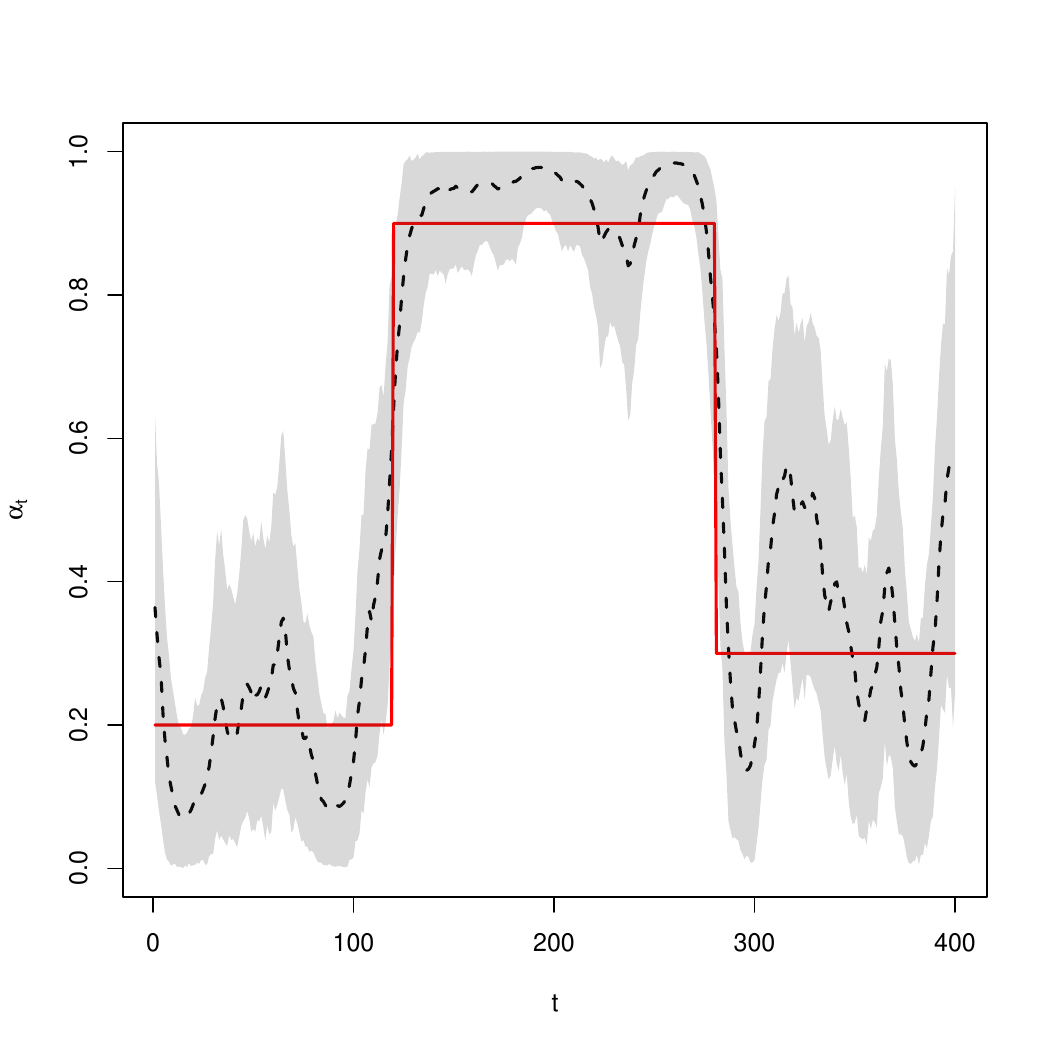}
\includegraphics[angle=0,width=0.3\linewidth]{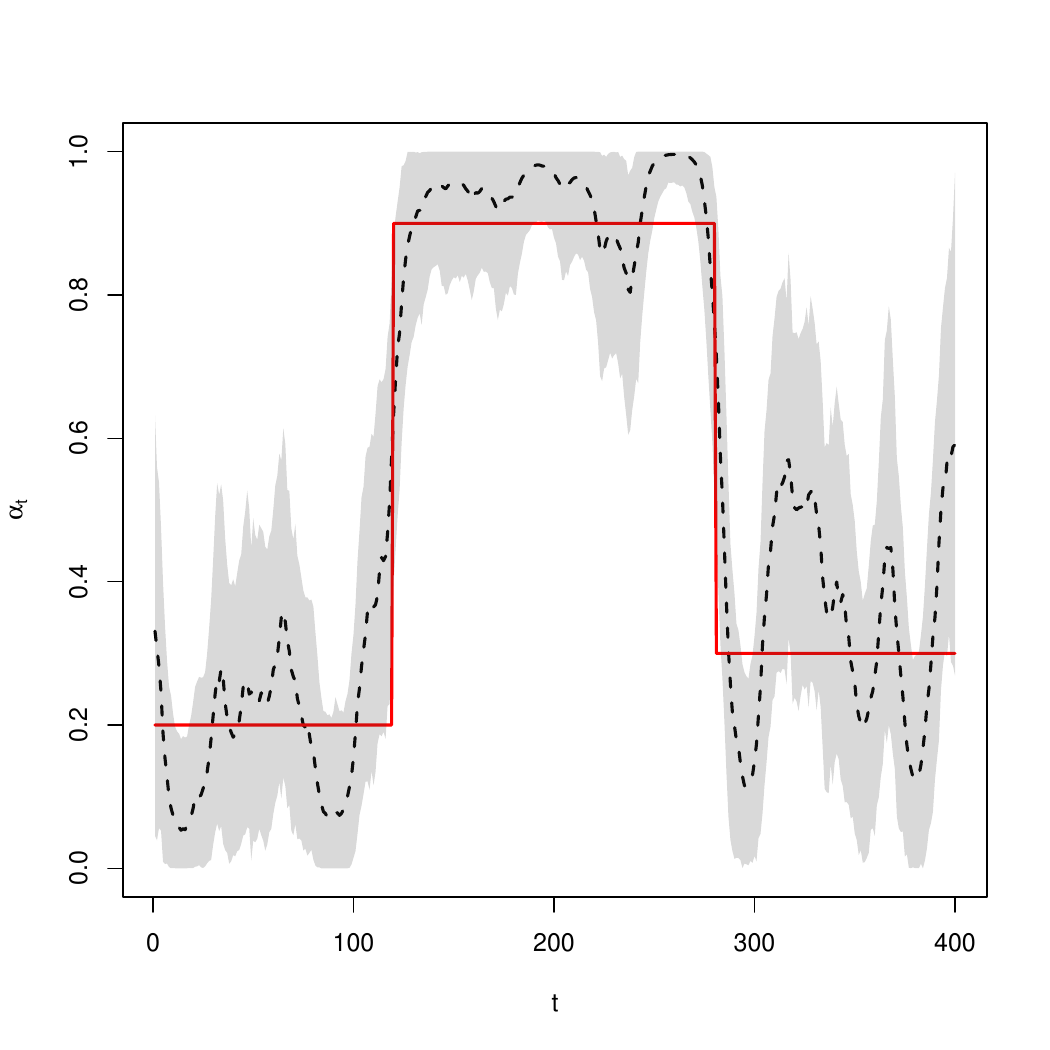}
\caption{Estimates (dashed lines) of the $ \alpha_t $'s (full lines) based on the mixture datasets. The mixture weights are $ \alpha_t^{(k)} $ ($ k $-th row), $ k = 1, 2, 3, 4 $. The first and second columns represent estimates based on the logit and probit transforms, respectively. The shaded area corresponds to the 90\% HPD intervals.}
\label{fig:mixfests}
\end{figure}

The estimates of the mixture weights are presented in Figure \ref{fig:mixfests}. One can see that the proposed method tends to provide good estimates, with shape that mimics the real curves. Furthermore, the estimates provided using probit and logit link functions are similar.

The performance of the method to estimate the component parameters is presented in Table \ref{tab:mixests}. One can see that the results using both link functions, logit and probit, are very similar. The method presents good point estimates, and most of the parameters belong to the 90\% HPD credible intervals. For the data generated using mixture weights with step behavior, the CI's of $ \mu_1 $ failed to  contain the true value (using both, logit and probit, link functions), as well as for $ \mu_2 $ in the case of using the logit. This happens most likely due to randomness. Although it is not presented here, we also estimated 95\% HPD credible intervals, where this issue was no longer observed. 
\begin{sidewaystable}\footnotesize
\centering
\caption{Estimates (medians and 90\% HPD credible intervals) for the component parameters $ \mu_1 = 0 $, $ \phi_1 = 4 $, $ \mu_2 = 2 $, $ \phi_1 = 4 $ of the mixture datasets generated according to each dynamical behavior: linear, parabolic, sinusoidal and steps. Both link functions, logit and probit, are considered.}
\label{tab:mixests}
\begin{tabular}{llrrrrrrrrrrrr}
\hline
&&\multicolumn{3}{c}{$ \mu_1 = 0 $} &\multicolumn{3}{c}{$ \phi_1 = 4 $} & \multicolumn{3}{c}{$ \mu_2 = 2 $} & \multicolumn{3}{c}{$ \phi_2 = 4 $} \\
\cline{3-5} \cline{6-8} \cline{9-11} \cline{12-14} 
Dynamic  & Link & Point & \multicolumn{2}{c}{90\% HPD CI} & Point & \multicolumn{2}{c}{90\% HPD CI} & Point & \multicolumn{2}{c}{90\% HPD CI} & Point & \multicolumn{2}{c}{90\% HPD CI} \\
Weight & & Estimate & Lower & Upper & Estimate & Lower & Upper & Estimate & Lower & Upper & Estimate & Lower & Upper \\ 
  \hline
\multirow{2}{*}{Linear} & Logit  & -0.030 & -0.091 & 0.038 & 4.081 & 3.274 & 4.979 & 1.965 & 1.900 & 2.039 & 4.093 & 3.184 & 4.905 \\ 
& Probit & -0.028 & -0.102 & 0.037 & 4.041 & 3.231 & 4.919 & 1.968 & 1.899 & 2.042 & 4.081 & 3.153 & 4.970 \\ 
\multirow{2}{*}{Parabolic} & Logit  & -0.018 & -0.071 & 0.034 & 4.245 & 3.546 & 4.958 & 2.059 & 1.997 & 2.130 & 4.380 & 3.418 & 5.511 \\ 
& Probit & -0.018 & -0.081 & 0.032 & 4.210 & 3.555 & 4.983 & 2.061 & 1.986 & 2.125 & 4.363 & 3.388 & 5.414 \\ 
\multirow{2}{*}{Sinusoidal} & Logit  & -0.022 & -0.086 & 0.041 & 4.140 & 3.343 & 5.039 & 2.008 & 1.956 & 2.065 & 4.765 & 3.943 & 5.679 \\ 
& Probit & -0.023 & -0.092 & 0.035 & 4.138 & 3.403 & 4.972 & 2.007 & 1.952 & 2.061 & 4.758 & 3.928 & 5.631 \\ 
\multirow{2}{*}{Steps} & Logit  & -0.072 & -0.143 & -0.004 & 4.499 & 3.516 & 5.631 & 1.924 & 1.851 & 1.992 & 3.557 & 2.843 & 4.284 \\ 
& Probit & -0.071 & -0.140 & -0.002 & 4.476 & 3.607 & 5.631 & 1.922 & 1.858 & 2.002 & 3.543 & 2.850 & 4.294 \\ 
   \hline
\end{tabular}
\end{sidewaystable}

\section{Application to the glioblastoma multiforme dataset}\label{sec:application}

The glioblastoma multiforme (GBM) dataset is related to a malignant tumor. The patient survival time for this kind of cancer has a median time of one year. The observations of the data are known as \emph{array Comparative Genomic Hybridization} (aCGH). They correspond to log-ratios of normalized intensities from disease \textit{vs.} control samples, which are indexed by the physical location of the probes on the genome \citep{LaiJohnsonKucherlapatiPark2005}. In other words, large values of aCGH suggest chromosomal aberrations in the specified locations. For this reason, the detection of regions with high proportions of abnormalities can be critical to comprehend the pathogenesis.

\begin{figure}
\centering
\includegraphics[angle=0,width=.5\linewidth]{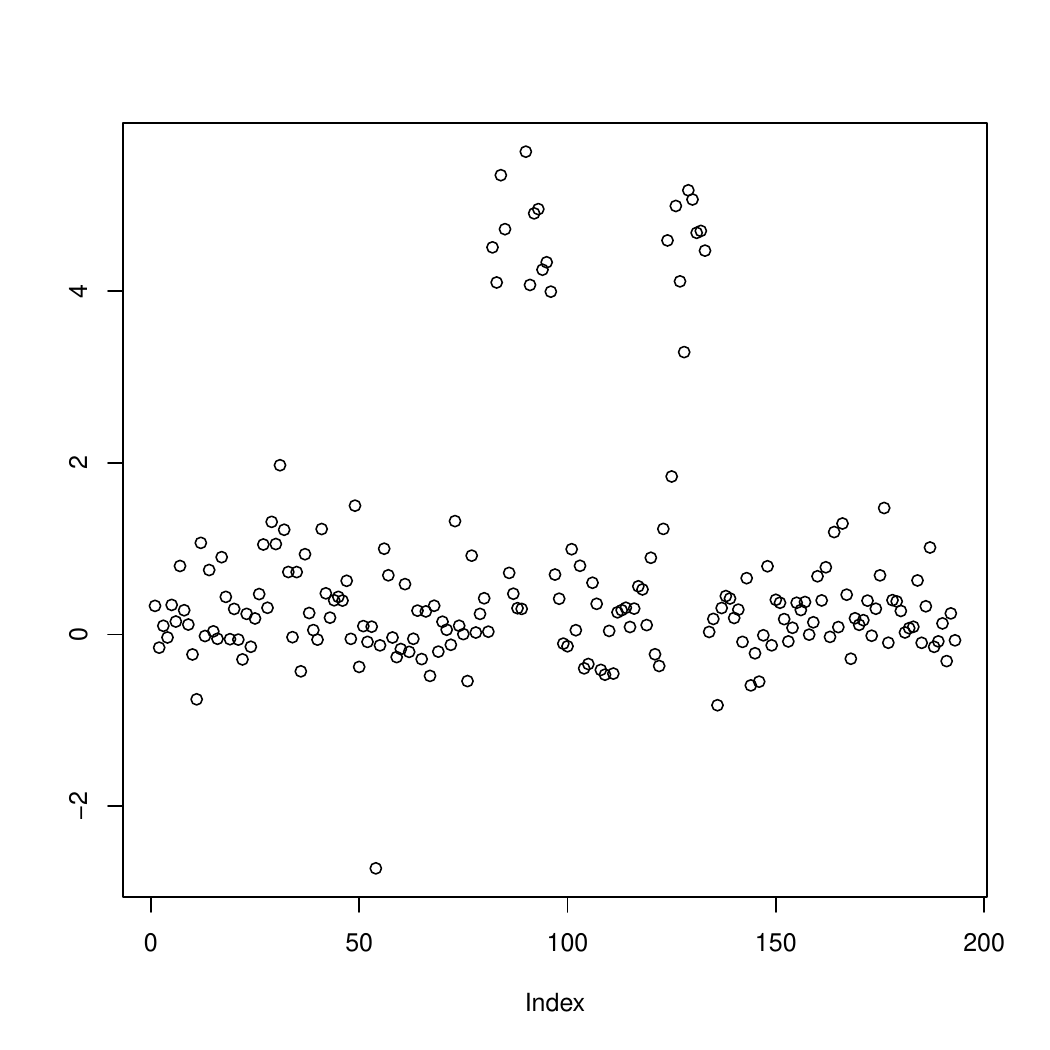}
\caption{Observed array Comparative Genomic Hybridization (aCGH) values. The values are log-ratios of normalized intensities from disease vs control samples, indexed by the physical location of the probes on the genome 
 \citep{LaiJohnsonKucherlapatiPark2005}.}
\label{fig:GBM}
\end{figure}

The data are presented in Figure \ref{fig:GBM}. They correspond to $ n = 193 $ aCGH observations. In this application we consider a mixture problem, where the observations can be treated as normal or aberrations. A similar study was performed by \cite{MontorilPinheiroVidakovic2019}. In their proposal, the authors needed to assume that the groups have known means, but in the application they had to estimate these parameters by averages and treat them as if they were the ``real'' ones. In the present paper, we estimate jointly both, the dynamic mixture weights and the mixture component parameters. Furthermore, credible intervals can also be provided, unlike in the aforementioned paper.

\begin{figure}
\centering
\includegraphics[angle=0,width=0.4\linewidth]{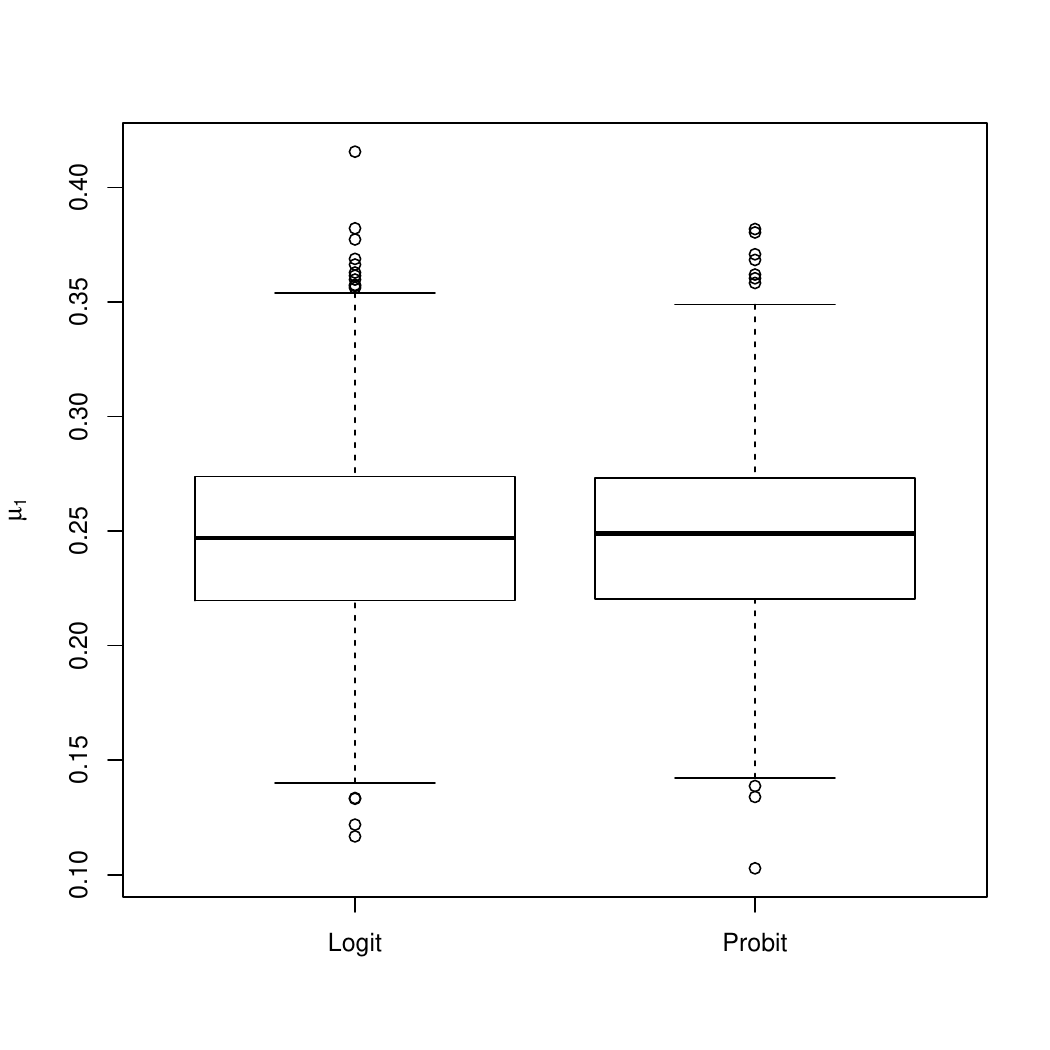}
\includegraphics[angle=0,width=0.4\linewidth]{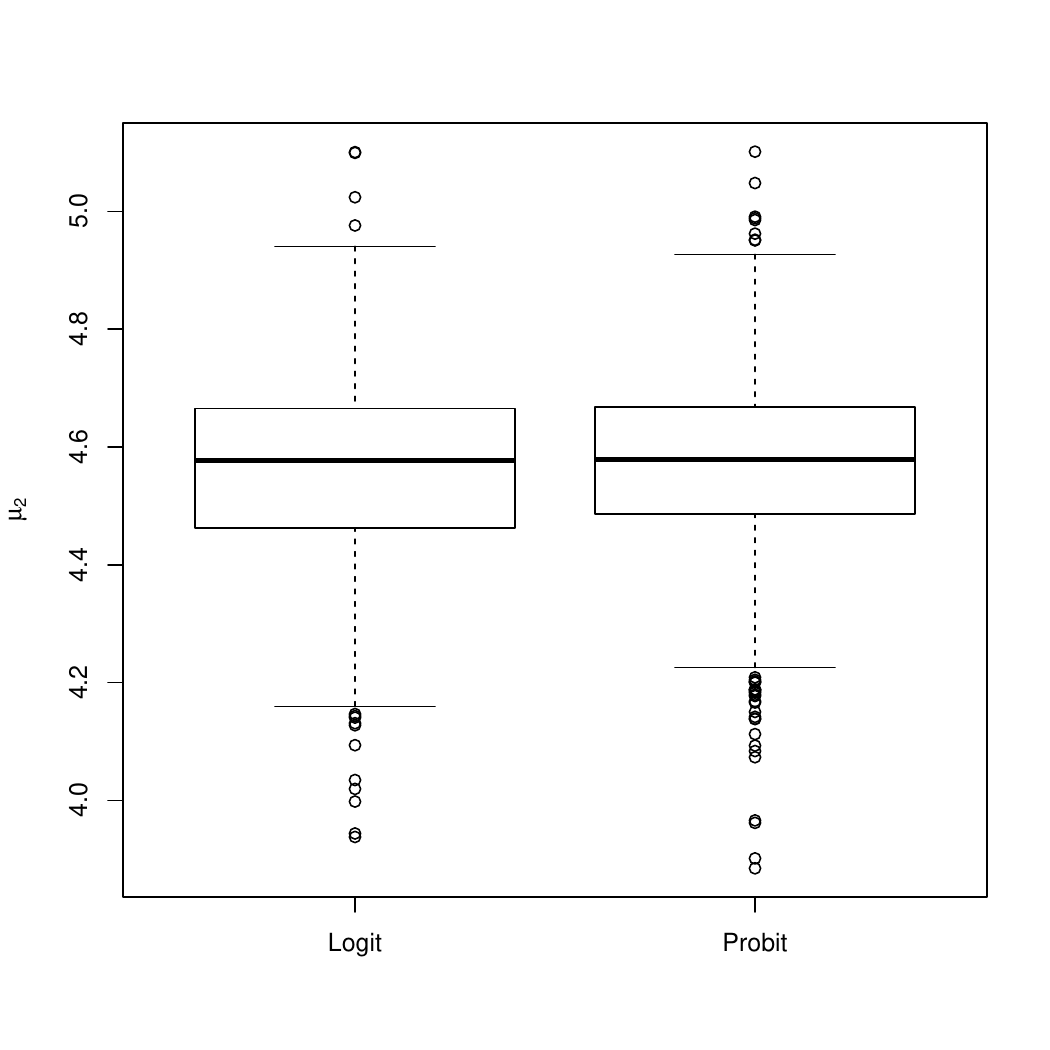} \\
\includegraphics[angle=0,width=0.4\linewidth]{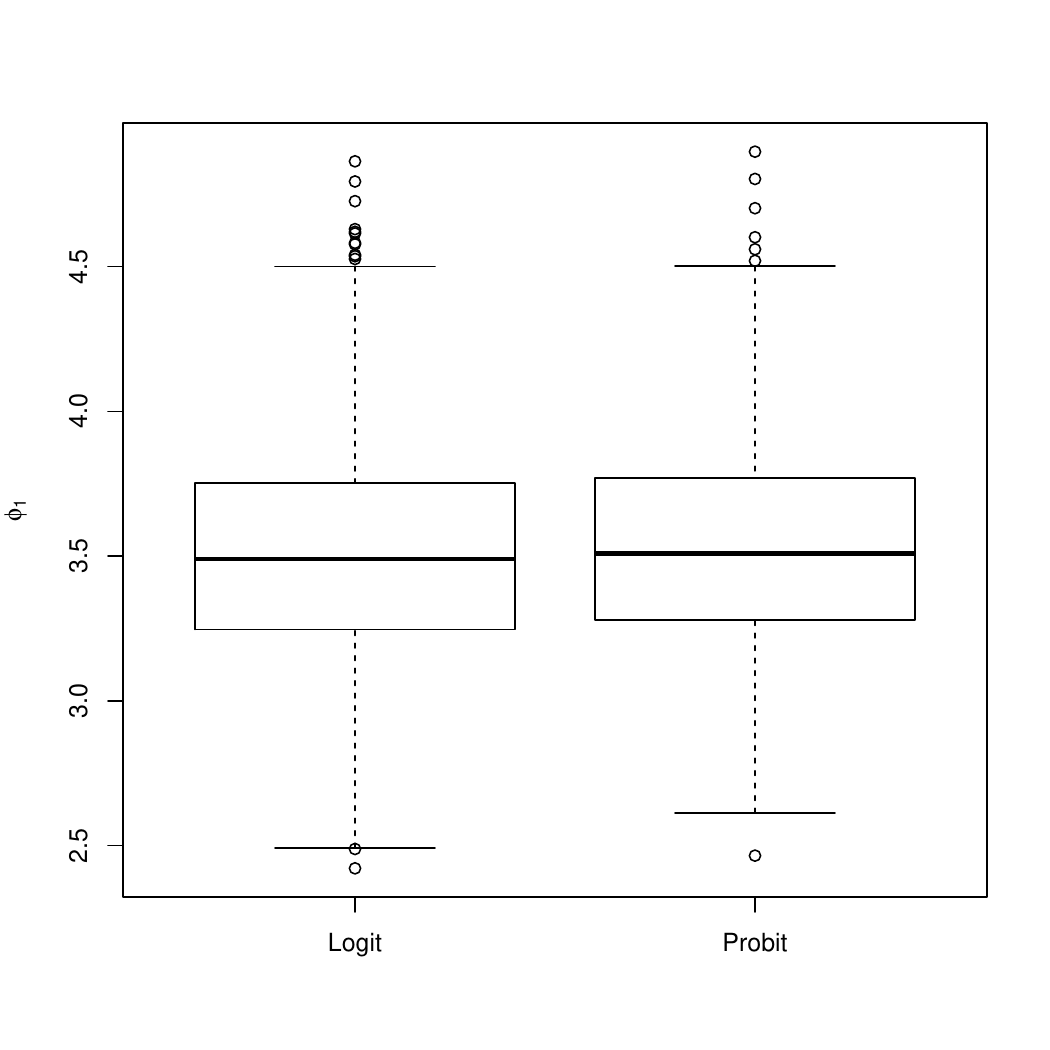}
\includegraphics[angle=0,width=0.4\linewidth]{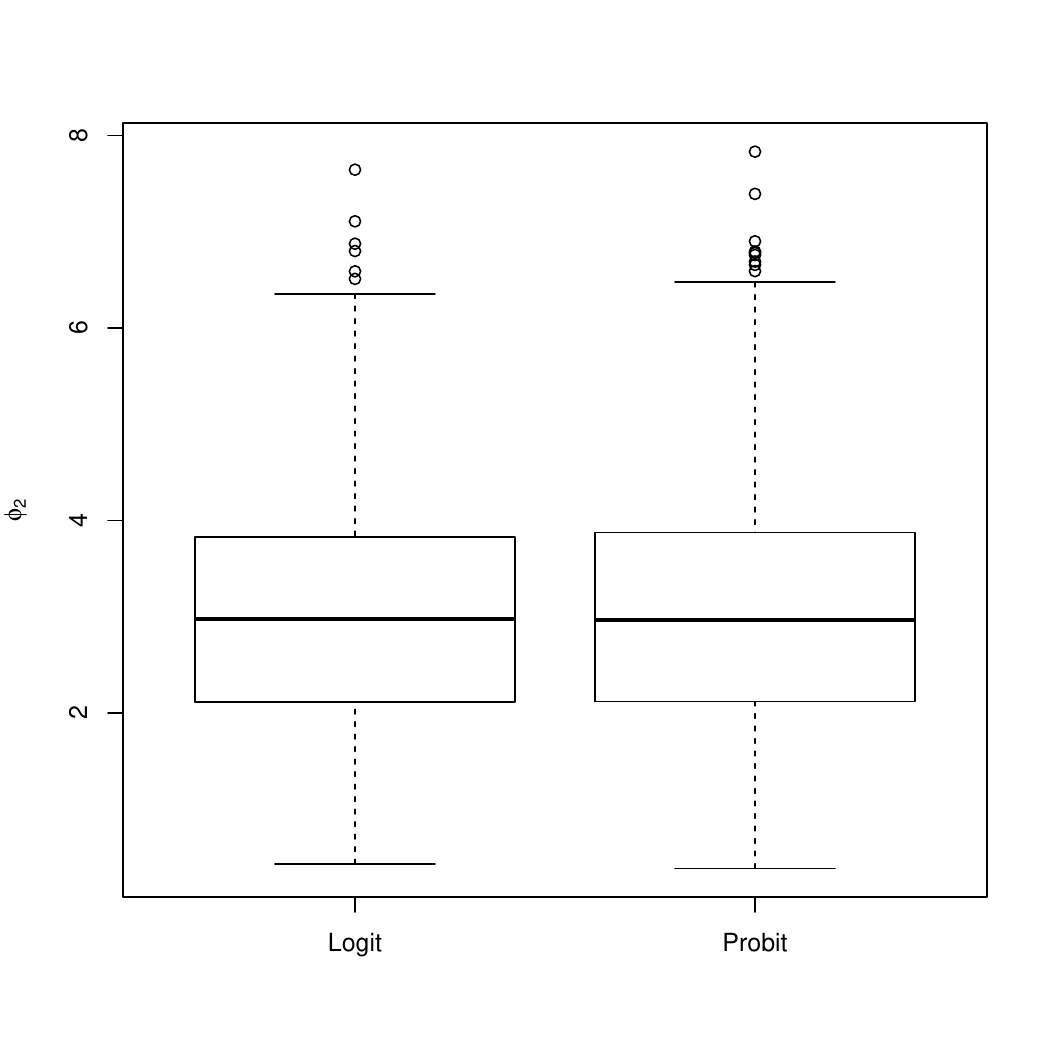}
\caption{Boxplots of the posterior of the component parameters generated via MCMC. Each graph presents the results based on the logit and probit transforms. The first and second rows represent means ($ \mu_1 $ and $ \mu_2 $) and precisions ($ \phi_1 $ and $ \phi_2 $), respectively.}
\label{fig:acgh-compparbp}
\end{figure}

In order to apply the proposed method to the data, we considered the same priors adopted in Section \ref{sec:mixdata}, and a second-order polynomial nonlinear dynamic model to estimate the mixture weights. We also adopted the same MCMC setup used in the previous section and we modeled the aCGH dataset using logit and probit link functions. Based on the MCMC results of the component parameters, estimates are summarized in Table \ref{tab:acgh-mixests}, where one can see similar results using both link functions. Also using these MCMC data, the behavior of the distribution of the posteriors is presented in Figure \ref{fig:acgh-compparbp}. The use of the logit or probit link function tends to provide posterior distributions that are very similar. A little more variability can be seen when the probit link function is adopted.
\begin{table}
\centering
\caption{Estimates (medians and 90\% HPD credible intervals) of the component parameters $ \mu_1 $, $ \phi_1 $, $ \mu_2 $, $ \phi_2 $ for the aCGH data. Both link functions, logit and probit, are considered.}
\label{tab:acgh-mixests}
\begin{tabular}{ccccccc}
\hline
&\multicolumn{3}{c}{Logit} & \multicolumn{3}{c}{Probit} \\
\cline{2-7}
Mixture & Point & \multicolumn{2}{c}{90\% HPD CI} & Point & \multicolumn{2}{c}{90\% HPD CI} \\
Parameter & Estimate & Lower & Upper & Estimate & Lower & Upper \\ 
  \hline
$ \mu_1 $ & 0.247 & 0.176 & 0.306 & 0.249 & 0.178 & 0.303 \\
$ \phi_1 $ & 3.491 & 2.903 & 4.143 & 3.510 & 2.929 & 4.180 \\
$ \mu_2 $ & 4.577 & 4.331 & 4.834 & 4.579 & 4.332 & 4.812 \\
$ \phi_2 $ & 2.977 & 1.046 & 4.773 & 2.965 & 0.992 & 4.894 \\
   \hline
\end{tabular}
\end{table}

\begin{figure*}
\centering
\includegraphics[angle=0,width=0.49\linewidth]{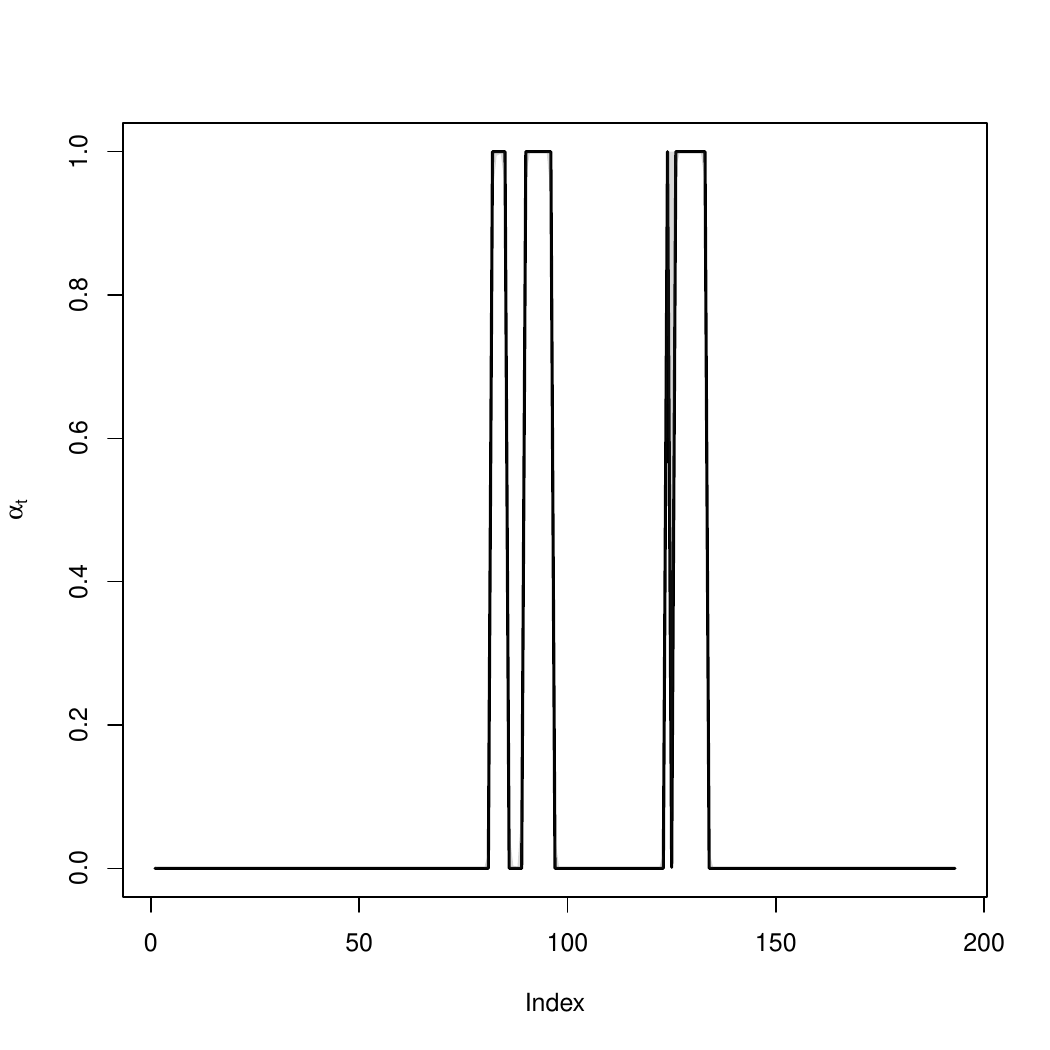}
\includegraphics[angle=0,width=0.49\linewidth]{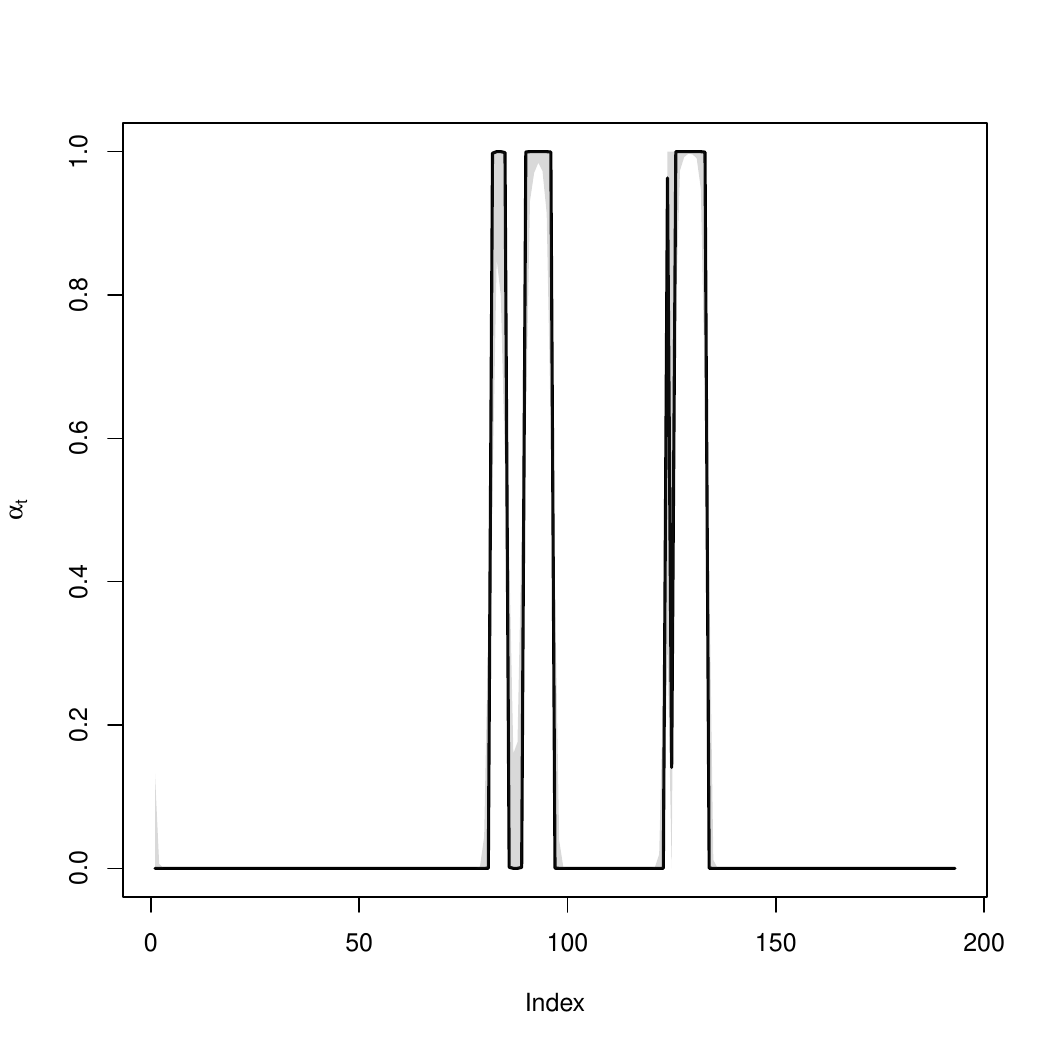}
\caption{Estimates (full lines) of the $ \alpha_t $'s the aCGH data. The top and bottom pictures represent estimates based on the logit and probit link functions, respectively. The shaded area corresponds to the 90\% HPD interval.}
\label{fig:acgh-mixfests}
\end{figure*}

With respect to the mixture weights, estimates are presented in Figure \ref{fig:acgh-mixfests}. Both HPD intervals are very tight, ensuring high precision for the point estimates. The use of the logit link function provides point estimates that suggest the existence of four regions with chromosome aberrations. Although it is not so easily seen, due to the tightness of the third peak and its proximity to the fourth, the 90\% HPD credible intervals reinforce the conclusion that there are four regions with chromosome aberrations (the lower limit for the third peak is around 0.8). When using the probit link function, the existence of four peaks is observed with high probability. These results presented are inline with the literature \citep[see][]{LaiJohnsonKucherlapatiPark2005}.

In comparison with \cite{MontorilPinheiroVidakovic2019}, 
the present proposal was able to detect the peaks with higher probabilities. Also, the third and fourth peaks presented here are in the same region as the third peak in the previous paper. The high probability presented by our method indicates the possibility of indices that can be better investigated.

The method is good to detect amplifications because the groups are separable. This makes it easier for the model to detect the groups clearly. Hence, the component parameter estimates tend to be unbiased, which in turn helps the estimation of the dynamic mixture weights.

\section{Conclusions and further remarks}\label{sec:comments}

In this work we propose, to the best of our knowledge, a new method to deal with Gaussian mixture models, where the mixture weights are allowed to have a dynamic behavior. The problem was studied with the use of polynomial dynamic models. We explored and developed properties for these models based on the ideas of \cite{Chan-Jeliazkov-2009}.

A general method, which can consider the estimation of the dynamic mixture weights as a particular case, was explored here, where two possibilities were approached, namely: (i) component-wise Metropolis-Hastings; and (ii) probit link function for Bernoulli data. In (ii), a probit link function was used to efficiently estimate dynamic curves. In (i), the method was able to use any (continuous and bijective) link function (we used the logit), although it was not as fast as (ii). In the simulation studies and in the application, both proposals provided similar results, with a little more variability of results based on (ii).

Due to the complexity of the problem and its wide applicability, we focused on the case of a dynamic mixture of two normal distributions. The general case encompasses $ K \geq 2 $ groups for the Gaussian mixture model and can be easily generalized. For example, one can use the precision-based algorithms discussed in Section \ref{sec:dinmixmodel} to estimate $ K $ independent curves $ \beta_{kt} $ such that $ \alpha_{kt} = \exp(\beta_{kt})/\sum_{k=1}^K \exp(\beta_{kt}) $, $ k = 1, 2, \ldots, K $, $ t = 1, 2, \ldots, T $. This transformation was used by \cite{Scaccia.Green-2003-JCGS}. A deeper analysis of such an extension will be left as topic of future research. Furthermore, although the dynamic mixture weights considered here are related only to the ``time'', it is also possible to include covariates to the model, without loss of efficiency to the methods of estimation proposed in Section \ref{sec:dinmixmodel}.

Another topic for future research is a scalable version of the dynamic generalized linear model, following \cite{West-Harrison-Migon-1985}. This is a combination of variational Bayes ideas with linear Bayes estimation. Two advantages of this approach are the recovery of sequential analysis, which allows for subjective intervention and faster processing time.

\section*{Acknowledgements}

The authors are grateful to Dr. Daiane A. Zuanetti, from the Department of Statistics at Federal University of S\~ao Carlos, Brazil, for the discussions and suggestions about mixture models.

\section*{Funding}

The first author was supported by FAPESP (Funda\c{c}\~{a}o de Amparo \`{a}
Pesquisa do Estado de S\~{a}o Paulo) Grant 2018/04654-9. The third author was supported by FAPESP Grant 1032, to visit the University of Campinas, and FAPERJ (Funda\c{c}\~{a}o de Amparo \`{a} Pesquisa do Estado do Rio de Janeiro) Grant E-26/007/10667/2019.

\bibliographystyle{tfnlm}
\bibliography{MyLibrary}

\end{document}



\title{Supplementary material of Bayesian estimation of dynamic weights in Gaussian mixture models}

\author{
\name{Michel H. Montoril\textsuperscript{a}\thanks{CONTACT M.~H. Montoril. Email: michel@ufscar.br}; Leandro T. Correia\textsuperscript{b} and Helio S. Migon\textsuperscript{c}}
\affil{\textsuperscript{a}Department of Statistics,
Federal University of S\~ao Carlos, S\~ao Carlos, Brazil; \\ \textsuperscript{b}Department of Statistics,
University of Bras\'ilia, Bras\'ilia, Brazil; \\
\textsuperscript{c}Department of Statistics,
Federal University of Rio de Janeiro, Rio de Janeiro, Brazil}
}

\maketitle

In this material we present a few proofs and derivations of results discussed in the paper. For the sake of clarity, we present the probability density function of normal and gamma \rv's in Section \ref{sec:pdfs}. The proofs are available in Section \ref{sec:proofs}
\appendix

\section{Main probability density functions}\label{sec:pdfs}
\subsection{Normal distribution}
\begin{definition}
We say that the random vector $ \vX = (X_1, \ldots, X_p)^{\T} $ has a normal distribution with mean $ \vmu $ and covariance matrix $ \Sigma $, i.e., $ \vX \sim N(\vmu, \Sigma) $, if its probability density function (pdf) can be written as
\[
p(\vx | \vmu, \Sigma) = \dfrac{1}{\sqrt{(2\pi)^p|\Sigma|}} \exp\ch{-\dfrac{1}{2} \pa{\vx - \vmu}^{\T} \Sigma^{-1} \pa{\vx - \vmu}}, \quad \vx \in \R^{p}.
\]
\end{definition}

The pdf above can be represented by its kernel, which corresponds to
\begin{equation}\label{kern:vnorm}
p(\vx | \vmu, \Sigma) \propto \exp\ch{ - \dfrac{1}{2} \vx^{\T} \Sigma^{-1} \vx +  \vmu^{\T} \Sigma^{-1} \vx}, \quad \vx \in \R^{p}.
\end{equation}

A usual and particular case corresponds to the situation where $ p = 1 $, that is, $ \vX $ is a random variable, and denoted by $ X $. In this case we have
\[
p(x | \mu, \sigma^2) = \dfrac{1}{\sqrt{2\pi \sigma^2}} \exp\ch{ -\dfrac{1}{2\sigma^2}(x-\mu)^2 }, \quad x \in \R.
\]
The kernel of the pdf above is
\begin{equation}\label{kern:norm}
p(x | \mu, \sigma^2) \propto \exp \ch{ -\dfrac{1}{2 \sigma^2} x^2 + \dfrac{\mu}{\sigma^2} x }, \quad x \in \R.
\end{equation}

\subsection{Gamma distribution}
\begin{definition}
We say that the random variable $ X $ has a gamma distribution with shape and rate parameters $ \nu  $ and $ \eta $, respectively, both positive, if its pdf is
\[
p(x | \nu, \eta) = \dfrac{\eta^{\nu}}{\Gamma(\nu)} x^{\nu - 1} \exp\ch{-\eta x}, \quad x > 0.
\]
One can write $ X \sim \Gamma(\nu, \eta) $.
\end{definition}
The kernel of the pdf above can be represented by
\begin{equation}\label{kern:gamma}
p(x | \nu, \eta) \propto x^{\nu - 1} \exp\ch{-\eta x}, \quad x > 0.
\end{equation}

\section{Proofs of some results presented in the manuscript}\label{sec:proofs}
\subsection{Modifying the precision-based algorithm in the polynomial case}
In this subsection, we intend to give more details about the main results presented in Section 2.1.1.

\subsubsection{Proof of Proposition 2.1}
In the manuscript we deal with a $ p $-th order polynomial DLM that can be written as
\begin{align}
y_t & = \theta_{t1} + \epsilon_t, \label{ssm:obs-eq} \\ 
\theta_{tk} & = \begin{cases}
\theta_{(t-1)k} + \theta_{(t-1)(k+1)} + \omega_{tk}, & k = 1, \ldots, p-1, \\
\theta_{(t-1)p} + \omega_{tp}, & k = p,
\end{cases} \label{ssm:st-eq}
\end{align}
where $ \epsilon_t \sim N(0, V) $ and $ \omega_{tk} \sim N(0, W_k) $, $ k = 1, \ldots, p $, $ t = 1, \ldots, T $, under the assumption that the innovations are independent. The model above makes clear the Markovian property (5) in the paper. Also, borrowing the matrix/vector notation used in the paper, it is easy to see that \eqref{ssm:obs-eq}-\eqref{ssm:st-eq} can be written as
\begin{align}
\vy & = \vvt_1 + \veps \label{ssm:obs-eq-mat} \\
\mH \vvt_k & = \begin{cases}
(\theta_{0k} + \theta_{0(k+1)}) \ve_1 + \mB \vvt_{k+1} + \vomega_k, & k = 1, \ldots, p-1, \\
\theta_{0p} \ve_1 + \vomega_{p}, & k = p,
\end{cases} \label{ssm:st-eq-mat}
\end{align}
where $ \mH $, the same in Proposition 2.1, can be written as

\begin{equation}\label{eq:mH}
\mH = \begin{pmatrix}
1 & 0 & 0 & \ldots & 0 & 0 \\
-1 & 1 & 0 & \ldots & 0 & 0 \\
0 & -1 & 1 & \ldots & 0 & 0 \\
\vdots & \vdots & \vdots & \ddots & \vdots & \vdots \\
0 & 0 & 0 & \ldots & -1 & 1
\end{pmatrix},
\end{equation}
$ \ve_1 = (1, 0, \ldots, 0)' $ and $ \mB = \mI - \mH $. After pre-multiplying both sides by $ \mH^{-1} $, one can see that
\begin{equation} 
\label{mod:mat-pol-ss}
\vvt_k = \vmu_k + \mH^{-1} \vomega_k, 
\end{equation}
where $ \vmu_k = (\theta_{0k} + \theta_{0(k+1)}) \ones + (\mH^{-1} - \mI) \vvt_{k+1} $, if $ k = 1, \ldots, p-1 $, and $ \vmu_p = \theta_{0p} \ones $. Therefore, \eqref{ssm:obs-eq-mat} and \eqref{mod:mat-pol-ss} yields the desired result in Proposition 2.1.

\subsubsection{Full conditional posterior of $ \vvt_k $ in (9)}
We derive the posterior for the case where $ k = 2, \ldots, p-1 $. For the case where $ k = 1 $ and $ k = p $, the result is obtained similarly. In order to simplify the notation, let us denote $ \mP_{k} = W_{k}^{-1} \mH^{\T}\mH $ and $ \mA = (\mH^{-1} - \mI) $. Therefore, due to Proposition 2.1, the posterior of the vector $ \vvt_{k} $ is determined by
\begin{align*}
p(\vvt_k | \fcond) & \propto p(\vvt_k | \vvt_{k+1}) p(\vvt_{k-1} | \vvt_k) \\
& \propto \exp\ch{ -\dfrac{1}{2} (\vvt_k - \vmu_{k})^{\T} \mP_{k} (\vvt_k - \vmu_{k}) } \\ & \qquad \times \exp\ch{ -\dfrac{1}{2} \pa{\vvt_{k-1} - \vmu_{k-1}}^{\T} \mP_{k-1} \pa{\vvt_{k-1} - \vmu_{k-1}}} \\
& \propto \exp\ch{ -\dfrac{1}{2} (\vvt_k - \vmu_{k})^{\T} \mP_{k} (\vvt_k - \vmu_{k}) } \\ & \quad \times \exp\ch{ -\dfrac{1}{2} \co{\vvt_{k-1} - (\theta_{0(k-1)} + \theta_{0k}) \ones - \mA \vvt_{k}}^{\T} \mP_{k-1} \co{\vvt_{k-1} - (\theta_{0(k-1)} + \theta_{0k}) \ones - \mA \vvt_{k}}} \\
& \propto \exp\bigg\{ -\dfrac{1}{2} \vvt_{k}^{\T} \mP_{k} \vvt_{k} + \vmu_{k}^{\T} \mP_{k} \vvt_{k} - \dfrac{1}{2} \vvt_{k}^{\T} \mA^{\T} \mP_{k-1} \mA \vvt_{k}  \\ & \qquad \qquad \quad + \pa{\vvt_{k-1} - (\theta_{0(k-1)} + \theta_{0k}) \ones}^{\T} \mP_{k-1} \mA \vvt_{k} \bigg\} \\
& \propto \exp \bigg\{ -\dfrac{1}{2} \vvt_{k}^{\T} \pa{ \mP_{k} + \mA^{\T} \mP_{k-1} \mA } \vvt_{k} \\ & \qquad\qquad\quad + \co{ \pa{\vvt_{k-1} - (\theta_{0(k-1)} + \theta_{0k}) \ones}^{\T} \mP_{k-1} \mA + \vmu_{k}^{\T} \mP_{k} } \vvt_{k} \bigg\} \\
& \propto \exp \bigg\{ -\dfrac{1}{2} \vvt_{k}^{\T} \pa{ \mP_{k} + \mA^{\T} \mP_{k-1} \mA } \vvt_{k} \\ & \qquad \qquad \quad + \co{ \mA^{\T} \mP_{k-1} \pa{\vvt_{k-1} - (\theta_{0(k-1)} + \theta_{0k}) \ones} + \mP_{k} \vmu_{k} }^{\T} \vvt_{k} \bigg\}.
\end{align*}

If we denote $ \bar{\mP}_{k} = \mP_{k} + \mA^{\T} \mP_{k-1} \mA $ and $ \bar{\vmu}_{k} = \bar{\Phi}_{k}^{-1} \co{ \mA^{\T} \mP_{k-1} \pa{\vvt_{k-1} - (\theta_{0(k-1)} + \theta_{0k}) \ones} + \mP_{k} \vmu_{k} } $, then
\begin{align*}
p(\vvt_{k} | \fcond) & \propto \exp\ch{ -\dfrac{1}{2} \vvt_{k}^{\T} \bar{\mP}_{k} \vvt_{k} + \bar{\vmu}_{k}^{\T} \bar{\mP}_{k} \vvt_{k} },
\end{align*}
which, by \eqref{kern:vnorm}, corresponds to the kernel of a multivariate normal distribution. Hence,
\[
\vvt_{k} | \fcond \sim N(\bar{\vmu}_{k}, \bar{\mP}_{k}^{-1}),
\]
where
\begin{align*}
\bar{\mP}_{k} & = \mA^{\T} \mP_{k-1} \mA + \mP_{k} = \frac{1}{W_{k-1}} \mA^{\T} \mH^{\T} \mH \mA + \frac{1}{W_{k}} \mH^{\T} \mH \\
& = \frac{1}{W_{k-1}} \mB^{\T} \mB + \frac{1}{W_{k}} \mH^{\T} \mH,
\end{align*}
with $ \mH\mA = \mH \pa{\mH^{-1} - \mI} = \pa{\mI - \mH} = \mB $, and
\begin{align*}
\bar{\vmu}_{k} & = \bar{\mP}_{k}^{-1} \ch{ \mA^{\T} \mP_{k-1} \co{\vvt_{k-1} - (\theta_{0(k-1)} + \theta_{0k}) \ones} + \mP_{k} \vmu_{k}} \\
& = \bar{\mP}_{k}^{-1} \bigg\{ \dfrac{1}{W_{k-1}} \mA^{\T} \mH^{\T} \mH \co{\vvt_{k-1} - (\theta_{0(k-1)} + \theta_{0k}) \ones} \\ & \qquad \qquad\quad + \dfrac{\theta_{0k} + \theta_{0(k+1)}}{W_{k}} \mH^{\T} \mH \ones + \mH^{\T} \mH \pa{\mH^{-1} - \mI} \vvt_{k+1} \bigg\} \\
& = \bar{\mP}_{k}^{-1} \ch{ \dfrac{1}{W_{k-1}} \mB^{\T} \mH \vvt_{k-1} + \dfrac{\theta_{0k} + \theta_{0(k+1)}}{W_{k}} \ve_1 + \dfrac{1}{W_{k}} \mH^{\T} \mB \vvt_{k+1} }.
\end{align*}
In the last equality above, we used two results. The first,
\[
\mH \ones = \begin{pmatrix}
1 & 0 & 0 & \ldots & 0 & 0 \\
-1 & 1 & 0 & \ldots & 0 & 0 \\
0 & -1 & 1 & \ldots & 0 & 0 \\
\vdots & \vdots & \vdots & \ddots & \vdots & \vdots \\
0 & 0 & 0 & \ldots & -1 & 1
\end{pmatrix} \begin{pmatrix}
1 \\
1 \\
1 \\
\vdots \\
1
\end{pmatrix} = \begin{pmatrix}
1 \\
0 \\
0 \\
\vdots \\
0
\end{pmatrix} = \ve_1,
\]
which provides $ \mH^{\T} \mH \ones = \ve_1 $. The second result is
\[
\mA^{\T} \mH^{\T} \mH \ones = \mA^{\T} \ve_1 = \begin{pmatrix}
0 & 1 & 1 & 1 & \ldots & 1 & 1 \\
0 & 0 & 1 & 1 & \ldots & 1 & 1 \\
0 & 0 & 0 & 1 & \ldots & 1 & 1 \\
\vdots & \vdots & \vdots & \vdots & \ddots & \vdots & \vdots \\
0 & 0 & 0 & 0 & \ldots & 0 & 1 \\
0 & 0 & 0 & 0 & \ldots & 0 & 0
\end{pmatrix} \begin{pmatrix}
1 \\
0 \\
0 \\
\vdots \\
0 \\
0
\end{pmatrix} = \begin{pmatrix}
0 \\
0 \\
0 \\
\vdots \\
0 \\
0
\end{pmatrix}.
\]

\subsubsection{Full conditional posterior of $ \theta_{0k} $ in (10)}
The full conditional posterior of $ \theta_{0k} $ can be derived using its prior and the distribution of the terms in the model where it is associated, which are
\begin{align*}
\theta_{1(k-1)} & = \theta_{0(k-1)} + \theta_{0k} + \omega_{1(k-1)}, \\
\theta_{1k} & = \theta_{0k} + \theta_{0(k+1)} + \omega_{1k}
\end{align*}
in \eqref{ssm:st-eq}. Thus, we have that

\begin{align*} 
p(\theta_{0k} | \fcond) & \propto p(\theta_{0k} | \mu_{\theta_{0k}}, \sigma_{\theta_{0k}}^2) p(\theta_{1(k-1)} | \theta_{0k}, \theta_{0(k-1)} W_{k-1}) p(\theta_{1k} | \theta_{0k}, \theta_{0(k+1)}, W_k) \\
& \propto \exp\ch{-\dfrac{1}{2\sigma_{\theta_{0k}}^2} (\theta_{0k} - \mu_{\theta_{0k}})^2} \exp\ch{-\dfrac{1}{2W_{k-1}} (\theta_{1(k-1)} - \theta_{0(k-1)} - \theta_{0k})^2} \\ & \qquad\qquad\quad \times \exp\ch{-\dfrac{1}{2W_{k-1}} (\theta_{1k} - \theta_{0k} - \theta_{0(k+1)})^2} \\
& \propto \exp\ch{ -\dfrac{1}{2 \sigma_{\theta_{0k}}^2} \theta_{0k}^2 + \dfrac{\mu_{\theta_{0k}}}{\sigma_{\theta_{0k}}^2} \theta_{0k} } 
\exp\ch{ -\dfrac{1}{2 W_{k-1}} \theta_{0k}^2 + \dfrac{\theta_{1(k-1)} - \theta_{0(k-1)}}{W_{k-1}} \theta_{0k} } \\ & \qquad\qquad\quad\times
\exp\ch{ -\dfrac{1}{2 W_{k}} \theta_{0k}^2 + \dfrac{\theta_{1k} - \theta_{0(k+1)}}{W_{k}} \theta_{0k} } \\
& \propto \exp\left\{ -\dfrac{1}{2} \pa{ \frac{1}{\sigma_{\theta_{0k}}^2} + \frac{1}{W_{k-1}} + \frac{1}{W_k} } \theta_{0k}^2 \right. \\ & \qquad\qquad\quad + \left. \pa{ \frac{\mu_{\theta_{0k}}}{\sigma_{\theta_{0k}}^2} + \frac{\theta_{1(k-1)} - \theta_{0(k-1)}}{W_{k-1}} + \frac{\theta_{1k} - \theta_{0(k+1)}}{W_{k}} } \theta_{0k} \right\} \\
& \propto \exp\left\{ \frac{1}{2 \pa{ \frac{1}{\sigma_{\theta_{0k}}^2} + \frac{1}{W_{k-1}} + \frac{1}{W_{k}} }^{-1} } \theta_{0k}^2 \right. \\ & \qquad\qquad\quad + \left. \dfrac{\pa{ \frac{1}{\sigma_{\theta_{0k}}^2} + \frac{1}{W_{k-1}} + \frac{1}{W_{k}} }^{-1} \pa{ \frac{\mu_{\theta_{0k}}}{\sigma_{\theta_{0k}}^2} + \frac{\theta_{1(k-1)} - \theta_{0(k-1)}}{W_{k-1}} + \frac{\theta_{1k} - \theta_{0(k+1)}}{W_{k}} }}{\pa{ \frac{1}{\sigma_{\theta_{0k}}^2} + \frac{1}{W_{k-1}} + \frac{1}{W_{k}} }^{-1}} \theta_{0k} \right\}.
\end{align*}
Thus, based on \eqref{kern:norm}, 
\[
\theta_{0k} | \fcond \sim N(\bar{\mu}_{\theta_{0k}}, \bar{\sigma}_{\theta_{0k}}^2),
\]
where
\begin{align*}
\bar{\sigma}_{\theta_{0k}}^2 & = \pa{ \frac{1}{\sigma_{\theta_{0k}}^2} + \frac{1}{W_{k-1}} + \frac{1}{W_{k}} }^{-1}, \\
\bar{\mu}_{\theta_{0k}} & = \bar{\sigma}_{\theta_{0k}}^2 \pa{ \frac{\mu_{\theta_{0k}}}{\sigma_{\theta_{0k}}^2} + \frac{\theta_{1(k-1)} - \theta_{0(k-1)}}{W_{k-1}} + \frac{\theta_{1k} - \theta_{0(k+1)}}{W_{k}} }.
\end{align*}

\subsubsection{Full conditional posterior of $ W_{k}^{-1} $ in (11) and $ V^{-1} $ in (12)}
The posterior distribution of the precision $ W_k^{-1} $ involves its prior distribution and the distribution of $ \vvt_{k}|\vvt_{k+1} $ in Proposition 2.1. Assuming that $ W_{k}^{-1} \sim \Gamma(\nu_{0k}, \eta_{0k}) $, we have
\begin{align*}
p(W_{k}^{-1}) & \propto p(W_{k}^{-1} | \nu_{0k}, \eta_{0k}) p(\vvt_{k} | \vvt_{k+1}) \\
& \propto \dfrac{1}{W_{k}^{\nu_{0k} - 1}} \exp\ch{-\dfrac{\eta_{0k}}{W_{k}}} \dfrac{1}{W_{k}^{T/2}}\exp\ch{-\dfrac{1}{2W_{k}} (\vvt_{k} - \vmu_{k})^{\T} \mH^{\T} \mH (\vvt_{k} - \vmu_{k})} \\
& \propto \pa{W_{k}^{-1}}^{\nu_{0k} + \frac{T}{2} - 1} \exp\ch{ -\co{\eta_{0k} + \frac{1}{2} (\vvt_{k} - \vmu_{k})^{\T} \mH^{\T} \mH (\vvt_{k} - \vmu_{k})} W_{k}^{-1} },
\end{align*}
which, by \eqref{kern:gamma}, ensures that
\[
W_{k}^{-1} | \fcond \sim \Gamma(\bar{\nu}_{0k}, \bar{\eta}_{0k}),
\]
where
\begin{align*}
\bar{\nu}_{0k} & = \nu_{0k} + \frac{T}{2}, \\
\bar{\eta}_{0k} & = \eta_{0k} + \frac{1}{2} (\vvt_{k} - \vmu_k)^{\T}\mH^{\T} \mH (\vvt_{k} - \vmu_k),
\end{align*}

The full conditional posterior of $ V^{-1} $ can be derived analogously.

\subsection{Proof of Theorem 2.2}
In order to prove this theorem, we make use of the following lemma (a sketch of its proof is provided at the end).

\begin{lemma}\label{lemma:fcond-beta}
Let $ \vb_{(k)} = (\beta_1, \beta_2, \ldots, \beta_{k-1}, \beta_{k+1}, \ldots, \beta_n)^{\T} $. If $ \vb \sim N(\vmu, W(\mH^{\T}\mH)^{-1}) $, with $ \mH $ defined as in \eqref{eq:mH}. Then,
\[
\beta_k|\vb_{(k)} \sim N(\tilde{\mu}_k, \tilde{\sigma}_k^2),
\]
where
\[
\tilde{\mu}_k = \begin{cases}
\mu_1 + \frac{1}{2}(\beta_2-\mu_2), & \text{if } k = 1, \\
\mu_k + \frac{1}{2} \co{(\beta_{k-1}+\beta_{k+1}) - (\mu_{k-1}+\mu_{k+1})}, & \text{if } k = 2, \ldots, n-1, \\
\mu_n + (\beta_{n-1}-\mu_{n-1}), & \text{if } k = n,
\end{cases}
\]
and
\[
\tilde{\sigma}_k^2 = \begin{cases}
\dfrac{W}{2}, & \text{if } k = 2, \ldots, n-1, \\
W, & \text{if } k = n.
\end{cases}
\]
\end{lemma}

Going back to the proof of Theorem 2.2, by (7) in Proposition 2.1,
\[ \vvt_{1} | \vvt_{2} \sim N\co{\vmu_1, W_1\pa{\mH^{\T}\mH}^{-1}}, \]
where $ \vmu_1 = (\theta_{01} + \theta_{02}) \ones + (\mH^{-1} - \mI) \vvt_{2} $. Then, by Lemma \ref{lemma:fcond-beta},
\[ \theta_{t1} | \vvt_{(t)1}, \vvt_{2} \sim N(\mu_{t1}^{*}, \tau_{t}^2), \]
where
\begin{equation}\label{eq:taut}
\tau_{t}^2 = \begin{cases}
\dfrac{W_1}{2}, & \text{if }  t = 1, \ldots, T - 1, \\
W_1, & \text{if } t = T.
\end{cases}
\end{equation}
With respect to the mean, one can see that
\[
\mH^{-1} - \mI = \begin{pmatrix}
0 & 1 & 1 & 1 & \ldots & 1 & 1 \\
0 & 0 & 1 & 1 & \ldots & 1 & 1 \\
0 & 0 & 0 & 1 & \ldots & 1 & 1 \\
\vdots & \vdots & \vdots & \vdots & \ddots & \vdots & \vdots \\
0 & 0 & 0 & 0 & \ldots & 0 & 1 \\
0 & 0 & 0 & 0 & \ldots & 0 & 0
\end{pmatrix}.
\]
Thus, each element of $ \vmu_1 $ can be written as
\begin{equation}\label{eq:mu1}
\mu_{t1} = \sum_{t=0}^{t-1} \theta_{t2} + \theta_{01} = \mu_{(t-1)1} + \theta_{(t-1)2},
\end{equation}
$ t = 1, \ldots, T $. Hence, based on \eqref{eq:mu1}, we have the following results. For $ t = 1 $,
\begin{align}\label{eq:mu11}
\begin{split}
\mu_{11}^{*} & = \mu_{11} + \dfrac{1}{2}(\theta_{21} - \mu_{12}) \\
& = \mu_{11} - \dfrac{1}{2} \mu_{21} + \dfrac{1}{2} \theta_{21} \\
& = \mu_{11} - \dfrac{1}{2} (\mu_{11} + \theta_{12}) + \dfrac{1}{2} \theta_{21} \\
& = \dfrac{1}{2} \mu_{11} + \dfrac{1}{2}(\theta_{21} - \theta_{12}) \\
& = \dfrac{1}{2} (\theta_{01} + \theta_{02}) + \dfrac{1}{2}(\theta_{21} - \theta_{12}).
\end{split}
\end{align}
For $ t = 2, \ldots, T - 1 $,
\begin{align}\label{eq:mu1t}
\begin{split}
\mu_{t1}^{*} & = \mu_{t1} + \dfrac{1}{2} \co{\pa{\theta_{(t-1)1} + \theta_{(t+1)1}} - \pa{\mu_{(t-1)1} + \mu_{(t+1)1}}} \\
& = \dfrac{1}{2} \co{\pa{\theta_{(t-1)1} + \theta_{(t+1)1}} + \pa{\mu_{t1} - \mu_{(t-1)1}} - \pa{\mu_{(t+1)1} - \mu_{t1}}} \\
& = \dfrac{1}{2} \pa{\theta_{(t-1)1} + \theta_{(t+1)1} + \theta_{(t-1)2} - \theta_{t2}} \\
& = \dfrac{1}{2} \pa{\theta_{(t+1)1} - \theta_{t2}} + \dfrac{1}{2} \pa{\theta_{(t-1)1} + \theta_{(t-1)2}}.
\end{split}
\end{align}
Finally, 
\begin{align}\label{eq:mu1T}
\begin{split}
\mu_{T1}^{*} & = \mu_{T1} + \pa{\theta_{(T-1)1} - \mu_{(T-1)1}} \\
& = \theta_{(T-1)2} + \theta_{(T-1)1}
\end{split}
\end{align}

The desired result follows from \eqref{eq:taut}, \eqref{eq:mu11}--\eqref{eq:mu1T}.

\begin{proof}{\bf of Lemma \ref{lemma:fcond-beta}.}
The algebra to prove this result is extremely tedious. Thus, for the sake of simplicity, let us consider the particular case where $ n = 5 $. We have that
\[
\mH^{\T} \mH = \begin{pmatrix}
2 & - 1 & 0 & 0 & 0 \\
-1 & 2 & - 1 & 0 & 0 \\
0 & -1 & 2 & - 1 & 0 \\
0 & 0 & -1 & 2 & - 1 \\
0 & 0 & 0 & -1 & 1 \\
\end{pmatrix} \quad \text{and} \quad
(\mH^{\T} \mH)^{-1} = \begin{pmatrix}
1 & 1 & 1 & 1 & 1 \\
1 & 2 & 2 & 2 & 2 \\
1 & 2 & 3 & 3 & 3 \\
1 & 2 & 3 & 4 & 4 \\
1 & 2 & 3 & 4 & 5 \\
\end{pmatrix}
\]
Let us consider $ k = 1 $. In this case, $ \vb $ is partitioned in $ \vb_1 = \beta_1 $ and $ \vb_2 = (\beta_2, \beta_3, \beta_4, \beta_5)^{\T} $. Denoting $ \vmu_1 = \mu_1 $, $ \vmu_2 = (\mu_2, \mu_3, \mu_4, \mu_5)^{\T} $, and $ \mS_{11} = W $, $ \mS_{12} = W (1, 1, 1, 1)^{\T} = \mS_{21}^{\T} $,
\[
\mS_{22} = W \begin{pmatrix}
2 & 2 & 2 & 2 \\
2 & 3 & 3 & 3 \\
2 & 3 & 4 & 4 \\
2 & 3 & 4 & 5 \\
\end{pmatrix},
\]
we can apply the well-known result to derive the conditional distribution of partitions of a multivariate normal vector. In other words, we have that 
\[ \beta_1 | \vb_{(1)} \sim N\pa{\tilde{\mu}, \tilde{\sigma}^2}, \]
where $\tilde{\sigma}^2 = \mS_{11} - \mS_{12} \mS_{22}^{-1} \mS_{21} = \frac{W}{2}$ and $$ \tilde{\mu} = \vmu_1 - \mS_{12} \mS_{22}^{-1} (\vb_2 - \vmu_2) = \mu_1 - (0.5, 0, 0, 0)(\vb_2 - \vmu_2) = \mu_1 - \frac{1}{2} (\beta_2 - \mu_2). $$
The remaining results for $ k = 2, 3, 4, 5 $ are derived analogously.
\end{proof}

\subsection{Full conditional posterior of $ \vvt_{1} $ in (13)}

In this case we have $ y_t | \alpha_t \sim \bern(\alpha_t) $, with $ \alpha_t = \Phi(\theta_{t1}) $ and $ \theta_{t1} $ representing the local level of the state equation, $ t = 1, \ldots, T $. Basically, one can write $ \alpha_t = P(y_t = 1) = \Phi(\theta_{t1}) \equiv \Phi(\ve_t^{\T} \vvt_{1}) $, where $ \ve_t $ corresponds to the $ t $-th column of an identity matrix of order $ T $. This identity matrix plays the role of the design matrix in the probit model. Therefore, the desired result comes from the data augmentation method proposed by \cite{Albert-Chib-1993}. In the present case $ \vvt_{1} $ has, by Proposition 2.1, the prior
\[ \vvt_{1}|\vvt_{2} \sim N\pa{\vmu_1, W_1 (\mH^{\T} \mH)}, \]
with $ \vmu_1 = (\theta_{01} + \theta_{02}) \ones + (\mH^{-1} - \mI) \vvt_{2} $.

\bibliographystyle{tfnlm} 
\bibliography{MyLibrary}   